\pdfoutput=1
\newcommand*{\ATLASLATEXPATH}{}
\documentclass[PAPER, cernpreprint, texlive=2016, UKenglish]{\ATLASLATEXPATH atlasdoc}
 
\usepackage[backend=bibtex]{\ATLASLATEXPATH atlaspackage}
\usepackage{\ATLASLATEXPATH atlasbiblatex}
 
\usepackage{\ATLASLATEXPATH atlasphysics}
 
\graphicspath{{logos/}{figures/}}
 
\usepackage{ANA-EXOT-2017-26-PAPER-defs}

\AtlasTitle{Search for heavy neutral leptons in decays of $W$ bosons produced in 13 TeV $pp$ collisions using prompt and displaced signatures with the ATLAS detector}

\AtlasAbstract{
The problems of neutrino masses, matter--antimatter asymmetry, and dark matter could be successfully addressed by postulating right-handed neutrinos with Majorana masses below the electroweak scale. In this work, leptonic decays of $W$ bosons extracted from 32.9 fb$^{-1}$ to 36.1 fb$^{-1}$ of 13 TeV proton--proton collisions at the LHC are used to search for heavy neutral leptons (HNLs) that are produced through mixing with muon or electron neutrinos. The search is conducted using the ATLAS detector in both prompt and displaced leptonic decay signatures. The prompt signature requires three leptons produced at the interaction point (either $\mu\mu e$ or $e e\mu$) with a veto on same-flavour opposite-charge topologies. The displaced signature comprises a prompt muon from the $W$ boson decay and the requirement of a dilepton vertex (either $\mu\mu$ or $\mu e$) displaced in the transverse plane by 4--300 mm from the interaction point. The search sets constraints on the HNL mixing to muon and electron neutrinos for HNL masses in the range 4.5--50 GeV.
}
 
\author{The ATLAS Collaboration}
 
\AtlasRefCode{EXOT-2017-26}
 
\PreprintIdNumber{CERN-EP-2019-071}

\AtlasJournal{JHEP}
\AtlasJournalRef{JHEP 10 (2019) 265}
\AtlasDOI{10.1007/JHEP10(2019)265}

\AtlasCoverSupportingNote{Search for short-lived heavy neutral leptons}{https://cds.cern.ch/record/2268148}
\AtlasCoverSupportingNote{Search for heavy neutral leptons produced in W boson decays using a displaced-vertex signature with the ATLAS detector}{https://cds.cern.ch/record/2267569}

\AtlasCoverAnalysisTeam{Ossama ABOUZEID, Geert-jan BESJES, Jyoti Prakash BISWAL, Mario CAMPANELLI, Saminder DHALIWAL, Arnaud DUBREUIL, Filips FEDOTOVS, Helle GORMSEN, Sergio GRANCAGNOLO, Heiko LACKER, Philippe MERMOD, Michele MICHETTI, Deywis MORENO, Gabriela NAVARRO, Kelsey O'CONNOR, Henriette Aarup PETERSEN, Carlos SANDOVAL USME, Craig SAWYER, Shohei SHIRABE, Abner SOFFER, Fabian THIELE, Stefania XELLA, Monika WIELERS}
 
\AtlasCoverEdBoardMember{Karsten K{\"O}NEKE~(chair), William BUTTINGER, Jochen DINGFELDER, Guillaume UNAL}

\AtlasCoverEgroupEditors{atlas-exot-2017-26-contact-editors@cern.ch}
 
\AtlasCoverEgroupEdBoard{atlas-EXOT-2017-26-editorial-board@cern.ch}
 
\hypersetup{pdftitle={ATLAS document},pdfauthor={The ATLAS Collaboration}}
 
\begin{document}
 
\newif\ifint
\intfalse
 
\maketitle
 
\tableofcontents
 
\section{Introduction}
\label{sec:intro}
 
The observation of neutrino oscillations implies that neutrinos are massive~\cite{Bellini2014}. This requires the addition of a neutrino mass-generation mechanism to the Standard Model (SM), which can manifest itself in the form of right-handed neutrinos, Majorana neutrinos, or both. Adding a right-handed Majorana neutrino (denoted by heavy neutral lepton HNL, or simply $N$) gives rise to the so-called Type-1 Seesaw mechanism~\cite{Yanagida1980}. An SM neutrino then acquires a mass inversely proportional to the HNL Majorana mass, providing a natural explanation for neutrino masses and why they are so small compared with those of other fermions. Heavy neutral leptons could generate the observed amount of baryon asymmetry in the universe through a process known as leptogenesis~\cite{Fukugita1986}, and an HNL with a mass of the order of keV would be a valid dark-matter candidate~\cite{Asaka2005}. In fact, the addition of three HNLs with masses below the electroweak scale, two of which are potentially accessible by accelerator-based experiments in the range $0.1$--$90$~GeV~\cite{Gorbunov2007}, could address the three fundamental questions of the origins of neutrino masses, baryon asymmetry, and dark matter~\cite{Asaka2005b,Canetti2013a}. Meeting these conditions requires small mixing angles between HNLs and neutrinos. Mixing requirements are relaxed if all three HNLs can participate in generating a baryon asymmetry~\cite{Canetti2014,Drewes2017}, which means however that none of the three HNLs is available as a dark-matter candidate.  Depending on the mixing and mass parameters, the HNL may decay promptly or be long-lived. In this paper, searches exploiting both prompt-decay and displaced-decay signatures are reported.
 
Heavy neutral leptons with masses below 5~GeV can be produced in hadron decays. In this case, an experimental strategy providing sensitivity to very low coupling strengths is to use a high-intensity beam on a fixed target and an instrumented decay volume at some distance from the production~\cite{Bernardi1988,CHARM1986,NuTeV1999,NA622017b,SHiP2015,SHiP2016}. Higher HNL masses can only be directly accessed through the decays of $W$, $Z$ or $H$ bosons, and indirectly through precision tests of the SM. Within some assumptions about the relative HNL mixing angles to the different neutrino flavours, experiments sensitive for processes such as $\mu\rightarrow e\gamma$ or $\mu\rightarrow eee$ can provide indirect constraints which are competitive with direct searches for HNL masses above 30~GeV~\cite{Antusch2015}. An analysis with the DELPHI experiment at LEP1 using $\sim$$10^6$ neutrinos from $Z$ boson decays provided the best direct constraints prior to the LHC in the HNL mass range $2$--$75$~GeV~\cite{Delphi1997}. At hadron colliders, HNLs are better sought in $W$ rather than $Z$ boson decays due to trigger requirements and the higher production cross section. The CMS Collaboration presented results~\cite{CMS2018} which explore HNL masses in the range $1$--$1200$~GeV and mixing to muon and electron neutrinos, using a signature of $W$ boson decays into three prompt leptons with lepton-number violation (LNV) similar to the prompt signature presented in this search. The high rate of $W$ boson production at the LHC, combined with the capability of the ATLAS and CMS experiments to reconstruct displaced vertices in their inner detectors, permits HNLs with longer lifetimes to be accessed in regions of the parameter space that are beyond the LEP constraints and allowed by cosmological constraints~\cite{Helo2014,Izaguirre2015,Mermod2017,Drewes:2015iva,Bertoni:2014mva}.

\begin{figure}[htbp]
\subfloat[]{\label{fig:feynman_mumue_lnc}
\includegraphics[width=0.45\textwidth]{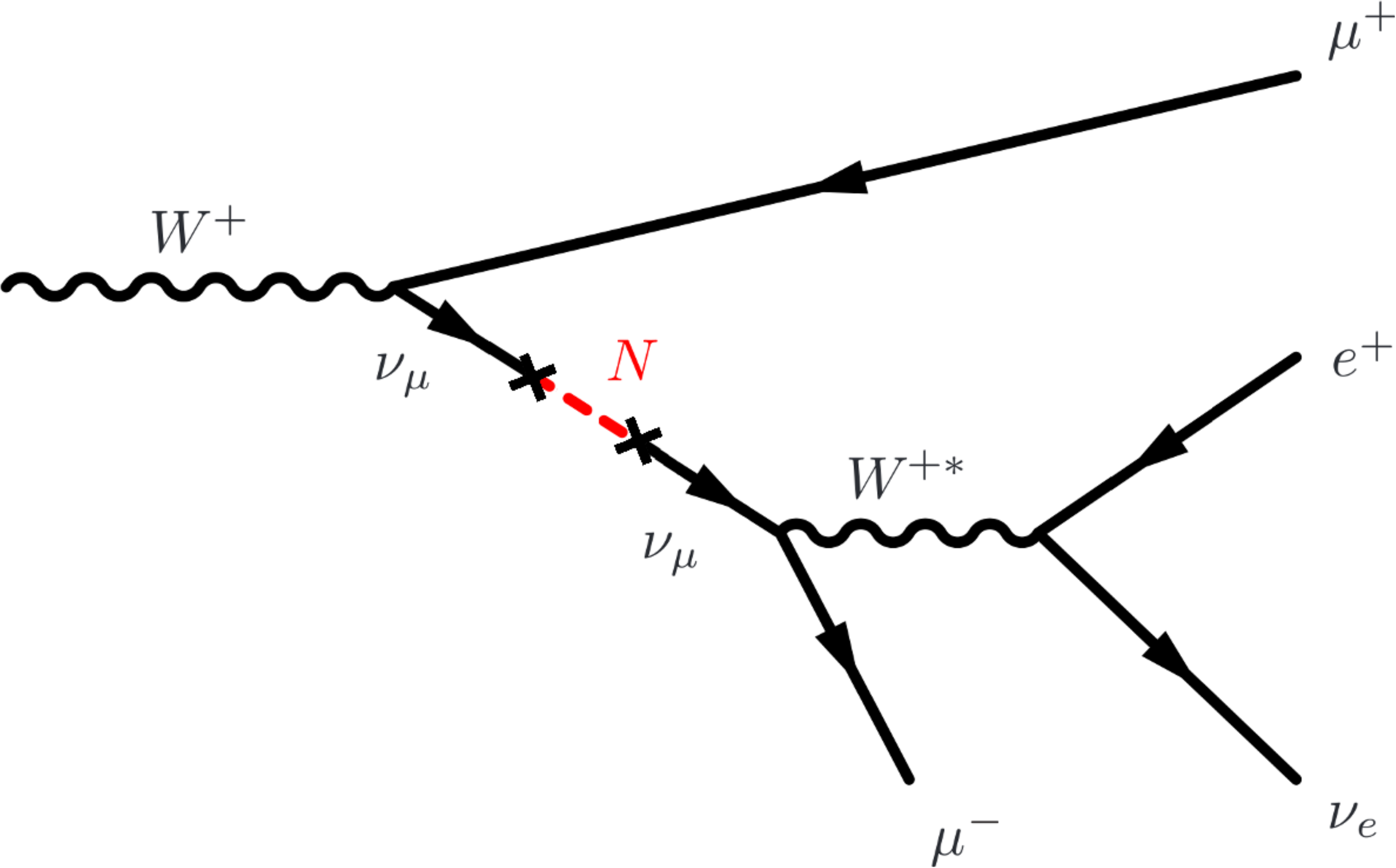}}
\hspace{0.05\textwidth}
\subfloat[]{\label{fig:feynman_mumumu_lnc}
\includegraphics[width=0.45\textwidth]{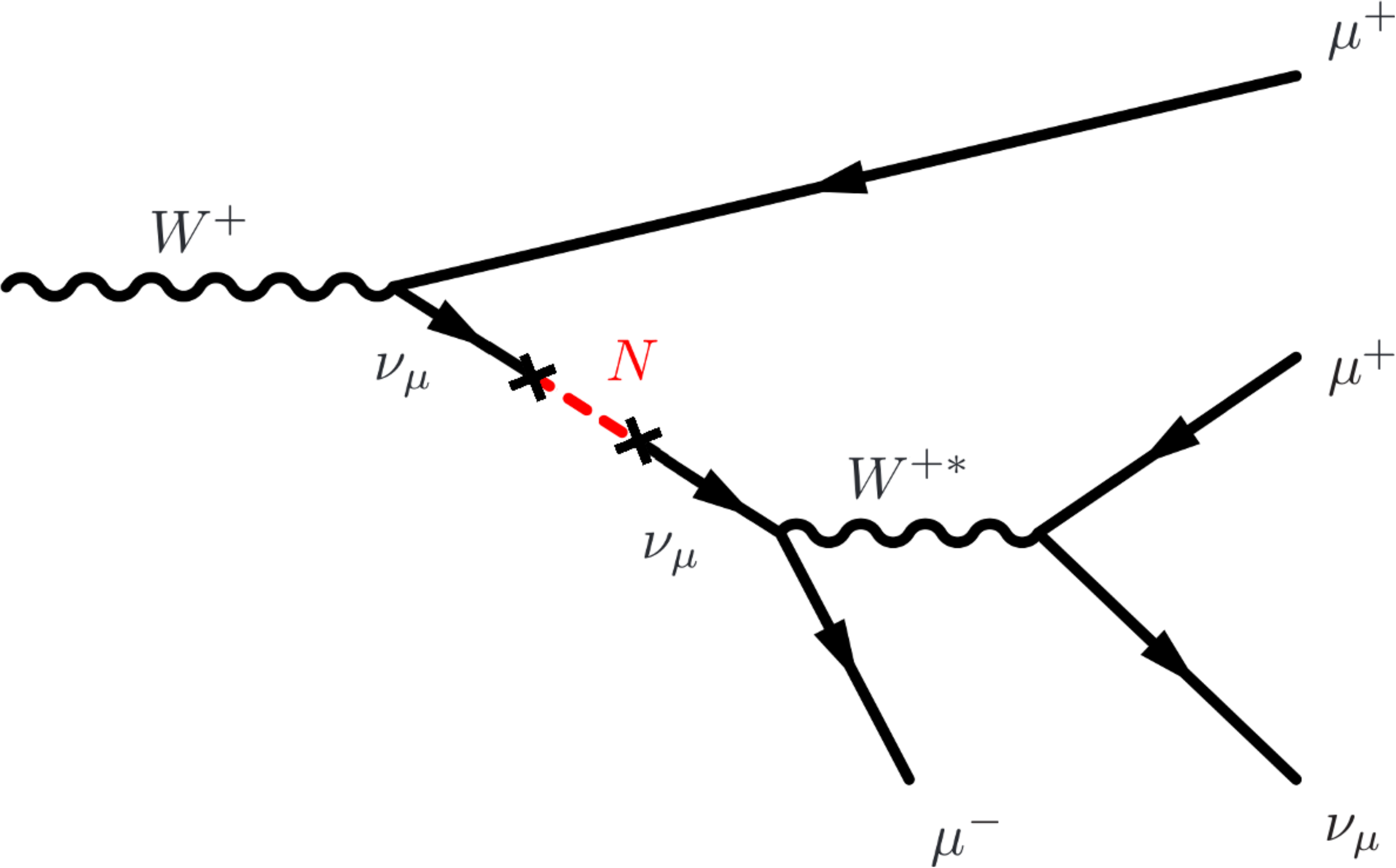}}
\newline
\subfloat[]{\label{fig:feynman_mumue_lnv}
\includegraphics[width=0.45\textwidth]{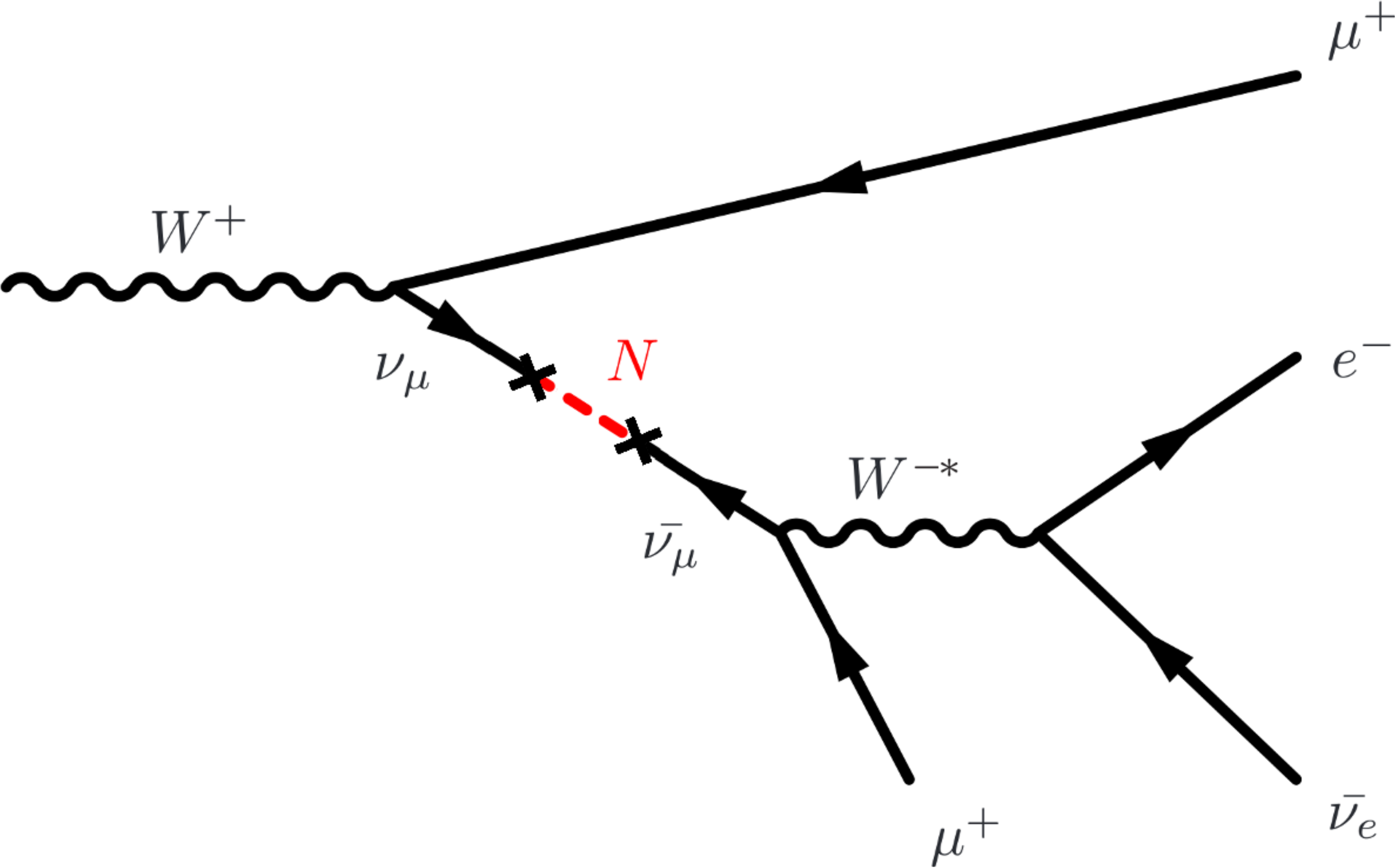}}
\hspace{0.05\textwidth}
\subfloat[]{\label{fig:feynman_mumumu_lnv}
\includegraphics[width=0.45\textwidth]{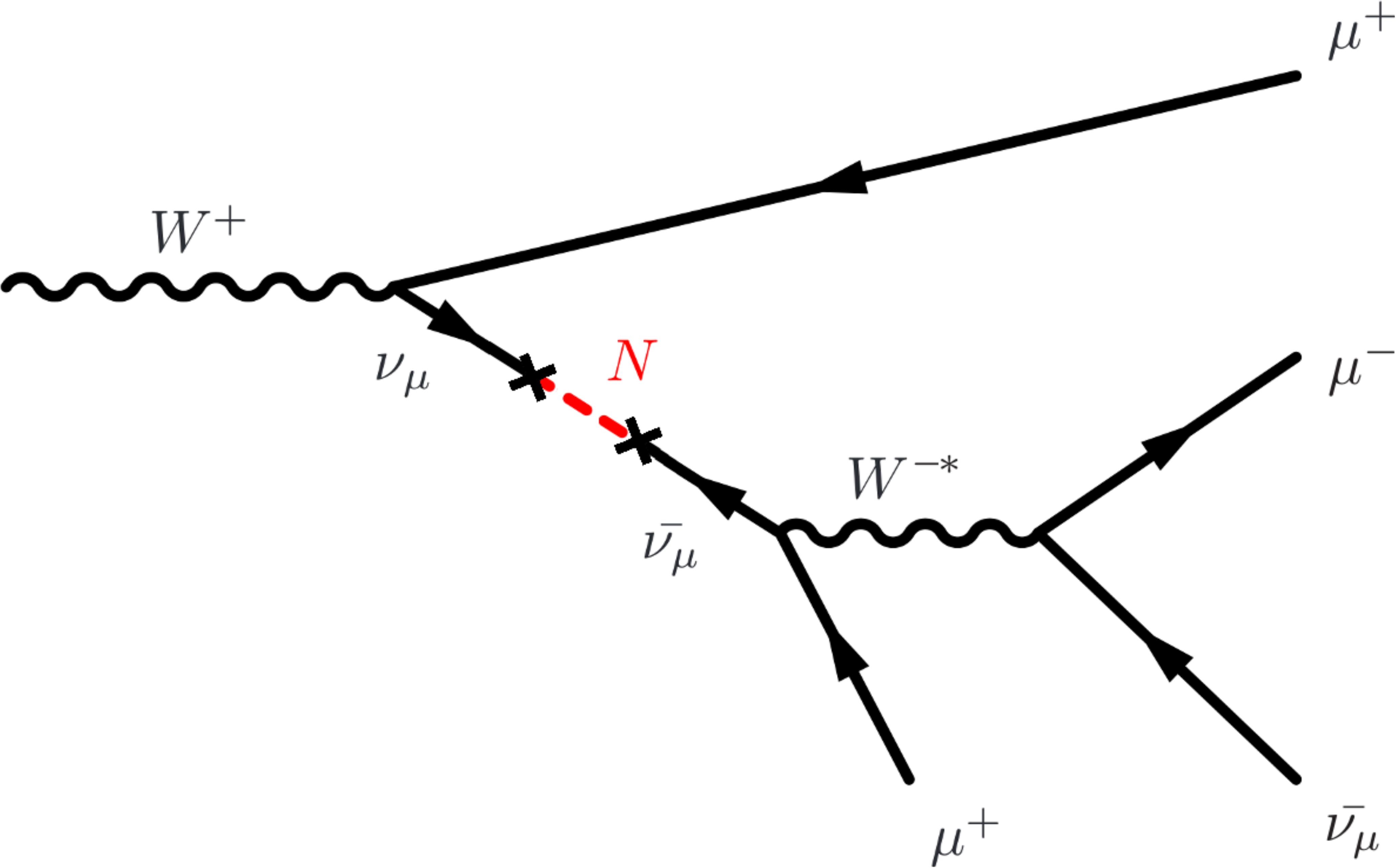}}
\hspace{0.05\textwidth}
\newline
\subfloat[]{\label{fig:feynman_eemu_lnv}
\includegraphics[width=0.45\textwidth]{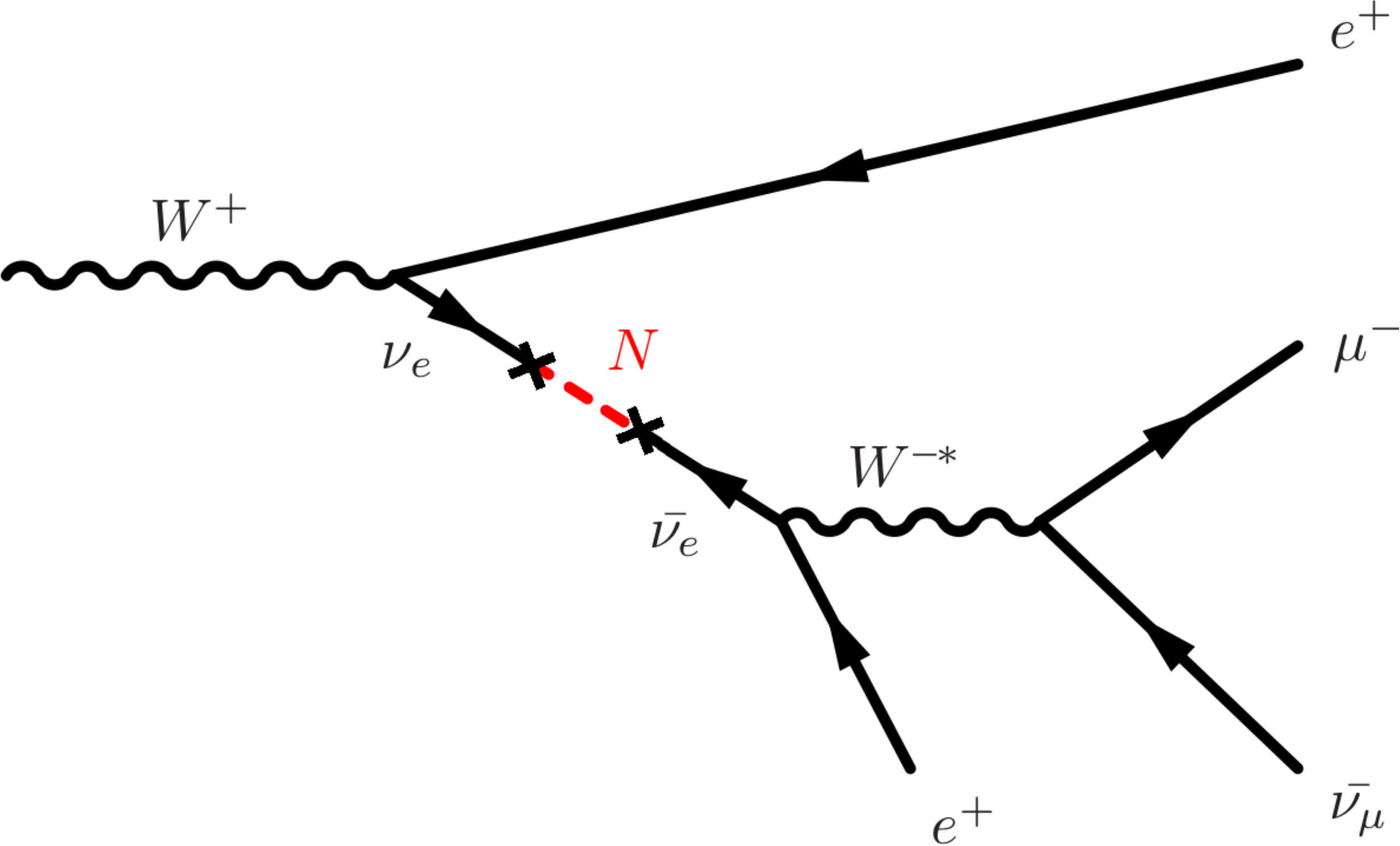}}
\caption{Feynman diagrams for $N$ production and decay in the channels which
this search is sensitive to: (a) $\mu$ mixing, $\mu e$ decay, LNC (probed by displaced signature); (b) $\mu$ mixing, $\mu\mu$ decay, LNC (probed by displaced signature); (c) $\mu$ mixing, $\mu e$ decay, LNV (probed by prompt and displaced signatures); (d) $\mu$ mixing, $\mu\mu$ decay, LNV (probed by displaced signature); (e) $e$ mixing, $e\mu$ decay, LNV (probed by prompt signature). Analogous processes involving the decay of the charge-conjugate $W^-$ boson are also included in the search, but omitted in this figure.}
\label{fig:feynman_diagrams}
\end{figure}
 
The present search relies on two distinct experimental signatures designed to probe both short and long HNL lifetimes. These select orthogonal sets of data taken in 2015--2016 with all ATLAS subsystems fully operational for proton--proton ($pp$) collisions at 13~TeV centre-of-mass energy. The dataset used in this analysis corresponds to an integrated luminosity of \SI{36.1}{\ifb} for the prompt signature and 32.9~fb$^{-1}$ for the displaced signature. The uncertainty in the integrated luminosity is \SI{2.1}{\percent}, derived from the calibration of the luminosity scale using $x$--$y$ beam-separation scans, following a methodology similar to that detailed in Ref.~\cite{ATLASlumi2016}, and using the LUCID-2 detector for the baseline luminosity measurements~\cite{Avoni2018}.
 
The prompt signature features three leptons originating from the interaction point, either two muons and an electron or two electrons and a muon, with same-flavour leptons of the same charge. This latter requirement is important for rejecting the large backgrounds from prompt SM processes. 
The displaced signature features a prompt muon accompanied by a vertex significantly displaced from the interaction point, formed by either two muons or a muon and an electron. The prompt lepton (expected to originate from a $W$ boson decay) is essential for ensuring an efficient triggering of such events and the displaced dilepton decay is very characteristic, rendering this signature virtually background-free. The selection of a prompt muon and a displaced leptonic decay is chosen for the first search of this type because requiring lepton objects (favouring muons as they are less readily affected by misidentification of other objects than electrons) considerably reduces QCD backgrounds, even though the same signature with a prompt electron and/or a displaced semileptonic decay can also be exploited at the LHC~\cite{Helo2014,Izaguirre2015,Mermod2017}.
 
The results of this search are interpreted in a model postulating a single right-handed Majorana neutrino $N$ produced in leptonic $W$ boson decays, with just two parameters: mass ($m_N$) and coupling strength ($|U|^2$). The heavy neutrino $N$ is allowed to mix exclusively with either $\nu_\mu$ or $\nu_e$ and to decay leptonically in a scenario with either lepton-number conservation (LNC) or LNV, as illustrated in Figure~\ref{fig:feynman_diagrams}. The Majorana nature of $N$ results in equal mixing to neutrinos and anti-neutrinos and the possibility for the process to violate the lepton number (although LNV is not guaranteed~\cite{Antusch2017}), as shown in Figures~\ref{fig:feynman_mumue_lnv},~\ref{fig:feynman_mumumu_lnv} and~\ref{fig:feynman_eemu_lnv}. The range $5<m_N<50$~GeV is explored using the prompt signature assuming LNV. The range $4.5<m_N<10$~GeV, corresponding to decay lengths of the order mm$-$cm, is probed down to lower $|U|^2$ values using the displaced signature with HNL mixing to $\nu_\mu$ without any assumption regarding LNC or LNV, as depicted in Figures~\ref{fig:feynman_mumue_lnc},~\ref{fig:feynman_mumumu_lnc},~\ref{fig:feynman_mumue_lnv} and~\ref{fig:feynman_mumumu_lnv}.

 
\section{HNL modelling}
\label{sec:model}
 
This section details $N$ production via mixing with an electron or muon neutrino originating from an on-shell $W$ boson decay, as well as its leptonic decays via the same mixing, as illustrated in Figure~\ref{fig:feynman_diagrams}. The generation and simulation of Monte Carlo (MC) signal and background events is presented at the end of the section.
 
\subsection{HNL production}
\label{sec:HNLprod}
 
The branching ratio of $W$ boson decays into a $N$ and a charged lepton, $\mathcal{B}(W\rightarrow \ell N)$, is proportional to the mixing matrix squared, or coupling strength, denoted by $|U|^2=\sum_{\ell} |U_{\ell}|^2$, where the terms $U_{\ell}$ are the matrix elements for $N$ mixing to the different neutrino flavours. The signatures considered in this search are sensitive to mixing to either $\nu_\mu$ or $\nu_e$ and can thus only constrain $|U_{\mu}|^2$ or $|U_e|^2$ (or potentially $|U|^2$ in the case where either of them is dominant).
 
The cross section times branching ratio for $W$ boson production and decay into $N$ and a charged lepton $\ell$ can then be expressed as~\cite{Dittmar1990}:
\begin{equation}
\label{equ:production}
\sigma(pp\rightarrow W)\cdot \mathcal{B}(W\rightarrow \ell N) =
\sigma(pp\rightarrow W)\cdot \mathcal{B}(W\rightarrow \ell\nu) \cdot
|U|^2\left(1-\frac{m_N^2}{m_W^2}\right)^2\left(1+\frac{m_N^2}{2m_W^2}\right).
\end{equation}
 
The product of the cross section for $W$ boson production in 13~TeV $pp$ collisions $\sigma(pp\rightarrow W)$ and the branching ratio for leptonic $W$ boson decay into a single lepton flavor $\mathcal{B}(W\rightarrow \ell\nu)$ (for $\ell=\mu$ or $e$) is taken from the ATLAS measurement in Ref.~\cite{ATLAS2016b} to be $20.6 \pm 0.6$~nb.
 
\subsection{HNL decay}
\label{sec:HNLdecay}
 
For this search, partial widths are calculated for all HNL decay channels including leptons and quarks. The calculations consider charged- and neutral-current-mediated interactions as well as QCD loop corrections, which are all described in Ref.~\cite{bondarenko_2018}. The HNL lifetime $\tau_N$ has a strong dependence on the coupling strength $|U|^2$ and also the mass $m_N$ due to phase-space effects. For a given $|U|^2$ and $m_N$, the total width $\Gamma = \sum_{i} \Gamma_i\left(m_N, |U|^2\right)$ is computed, and the mean lifetime is obtained as $\tau_N = \hbar/\Gamma$. In the relevant range $4.5\leq m_N\leq 50$~GeV, the result agrees within 2\% with the following parameterisations given in Ref.~\cite{Gronau1984}:  $\tau_{N_\mu} = (4.49\cdot 10^{-12}~\text{s})|U|^{-2}(m_N/1~\text{GeV})^{-5.19}$ and $\tau_{N_{e}} = (4.15\cdot 10^{-12}~\text{s})|U|^{-2}(m_N/1~\text{GeV})^{-5.17}$ for dominant mixing to $\nu_\mu$ and $\nu_e$, respectively. These relationships, however, assume no LNV decays. If LNV is allowed, twice as many decay channels are allowed, and $\tau_N$ is reduced by a factor of 2. More elaborate models do not necessarily allow for LNV~\cite{Antusch2017} and thus may or may not contain this factor of 2. To account for this model dependence, both interpretations are considered in the case of the displaced signature, which is not limited to LNV processes.
 
Leptonic HNL decay branching ratios are determined from the partial decay widths relative to the total width. In the mass range 4.5--50~GeV they have almost no mass dependence and yield $\mathcal{B}(N\rightarrow \ell\ell\nu_
\ell)=0.060$ and $\mathcal{B}(N\rightarrow \ell \ell' \nu_{\ell'})=0.106$ for dominant mixing to a given lepton species $\ell = \mu \text{ or } e$ ($\ell \neq \ell'$, including both charges). The difference between decays into leptons of the same flavour and different flavour is due to interference between decays through $W$ and $Z$ boson mediators, which is only present in the same-flavour case. This calculation and calculations found in the literature~\cite{Gronau1984,Gorbunov2007,Atre2009} can yield up to 5\% relative differences, mainly due to different treatments of QCD corrections. This 5\% difference is considered as a relative theoretical systematic uncertainty in the branching ratio.
 
\subsection{Event generation and simulation}
\label{sec:MC}
 
The signal MC samples were generated with a model of $W$ boson production in
13~TeV $pp$ collisions, with the $W$ exclusively decaying into a muon or electron
and an HNL ($W\rightarrow \mu N$ or $W\rightarrow e N$). Separate signal samples are generated for multiple choices of the HNL mass, $m_N=$~\SIlist[list-final-separator =
{~or~}]{4.5;5;7.5;10;12.5;15;20;30;50}{\GeV}, and the mean proper decay length, $c\tau=$~\SIlist[list-final-separator =
{~or~}]{0.001;0.01;0.1;1;10;100}{\milli\meter}, with $\sim 50000$ events per sample. The $N$ decay modes correspond to the
diagrams shown in Figure~\ref{fig:feynman_diagrams}. This model is implemented in \textsc{Pythia8}~\cite{sjostrand_introduction_2015} (v8.210), using the
NNPDF2.3 LO~\cite{Ball2013} parton distribution function (PDF) set and the A14 set of tuned parameters (tune)~\cite{ATLAS2012d}.
 
Background processes were generated using
\textsc{Powheg-Box}~\cite{alioli_general_2010,nason_new_2004,frixione_matching_2007}
with the next-to-leading order (NLO) CT10 PDF set~\cite{lai_new_2010} for top-quark pair ($t\bar{t}$) (using \textsc{v2} in r3026) and single top-quark (using r2856) production, in
combination with \textsc{Pythia}~\cite{sjostrand_pythia_2006} (v6.428, for parton showering) using the
CTEQ6L1 PDF set~\cite{pumplin_new_2002} and Perugia 2012 tune~\cite{skands_tuning_2010}.
\textsc{MadGraph5\_aMC@NLO}~\cite{alwall_automated_2014} (v2.2.2.p6) with the
NNPDF2.3 LO PDF set and A14 tune was used for $t\bar{t}W$ and $t\bar{t}Z$, while
\textsc{Sherpa}~\cite{gleisberg_event_2008} (v2.2.1) with the NNPDF2.3 LO PDF set was used for $WW$,
$ZZ$, $WZ$, $W$+jets, $Z$+jets and tribosons. The purely leptonic decays of dibosons
were generated using \textsc{Powheg-Box v2} (r2819) with NLO CT10 PDF set in
combination with \textsc{Pythia8} using the
CTEQ6L1 PDF set with AZNLO tune~\cite{ATL-PHYS-PUB-2013-017}. Together with the
\textsc{Pythia8}, \textsc{Powheg-Box} and \textsc{MadGraph5} event generators, \textsc{EvtGen}~\cite{Lange2001} (v1.2.0) was used for $b$- and $c$-hadron decays.
Also, together with \textsc{Powheg-Box} in processes involving $t$-quarks, \textsc{Tauola}~\cite{chrzaszcz_tauola_2016} (v27feb06) was used for $\tau$-lepton decays and \textsc{Photos}~\cite{golonka_photos_2006} (v2.09) was used for QED corrections.
 
For the generation of $t\bar{t}$ events, the top-quark mass was set to 172.5
GeV. The sample is normalised using the next-to-next-to-leading-order (NNLO)
cross section, including soft-gluon resummation to
next-to-next-to-leading-logarithm (NNLL) \cite{beneke_hadronic_2012, cacciari_top-pair_2012, baernreuther_percent_2012, czakon_nnlo_2012, czakon_nnlo_2013, czakon_total_2013, czakon_top++:_2014}.
For events containing a $W$ or $Z$ boson with associated jets simulated using \textsc{Sherpa}, matrix elements were generated with up to two additional parton emissions at NLO accuracy and up to four additional parton emissions at LO accuracy using \textsc{COMIX}~\cite{Gleisberg:2008fv} and \textsc{OpenLoops}~\cite{Cascioli:2011va} and merged with the \textsc{Sherpa} parton shower~\cite{Schumann:2007mg} according to the ME+PS@NLO prescription~\cite{Hoeche:2012yf}. Diboson processes with one of the bosons decaying hadronically and the other leptonically are calculated for up to one ($ZZ$) or zero ($WW$, $WZ$) additional partons at NLO and up to three additional partons at LO using \textsc{COMIX} and \textsc{OpenLoops}, and merged with the \textsc{Sherpa} parton shower according to the ME+PS@NLO prescription. Matrix elements of triboson processes containing two or more charged leptons were generated with \textsc{Sherpa} including off-shell contributions with up to one additional real emission at LO accuracy~\cite{ATL-PHYS-PUB-2017-005}.
 
The generated events were processed through a full simulation of the ATLAS detector geometry and response~\cite{ATLAS2010} using the Geant4~\cite{Geant42002} toolkit.
The simulation includes multiple $pp$ interactions per bunch crossing (pile-up), as well as the effect on the detector response due to interactions from bunch crossings before or after the one containing the hard interaction. Pile-up was simulated with the soft processes of \textsc{Pythia8} using the A2 tune~\cite{ATLAS2012d} and the MSTW2008LO PDF~\cite{Martin2009}. Per-event weights were applied to the simulated events to reproduce the distribution of the average number of interactions per bunch-crossing as measured in data.

 
\section{The ATLAS detector}
\label{sec:detector}

The ATLAS experiment~\cite{PERF-2007-01,ATLAS-TDR-19,ATLAS2018f} at the LHC is a
multipurpose particle detector with a forward--backward symmetric
cylindrical geometry and a nearly $4\pi$ coverage in solid
angle.\footnote{ATLAS uses a right-handed coordinate system with its
origin at the nominal IP in the centre of the detector and the
$z$-axis along the beam pipe. The $x$-axis points from the IP to the
centre of the LHC ring, and the $y$-axis points upward. Cylindrical
coordinates $(r,\phi)$ are used in the transverse plane, $\phi$
being the azimuthal angle around the $z$-axis. The pseudorapidity
is defined in terms of the polar angle $\theta$ as
$\eta=-\ln\tan(\theta/2)$. Angular distance is measured in units of $\Delta R = \sqrt{(\Delta \eta)^2 + (\Delta \phi)^2}$.}
The detector consists of several layers
of subdetectors. From the interaction point (IP) outwards there is an inner tracking
detector (ID), electromagnetic and hadronic calorimeters, and a muon
spectrometer (MS).
 
The ID extends from a cylindrical radius of about 33 to 1100~mm and
to $|z|$ of about 3100~mm, and is immersed in a 2~T axial magnetic
field. It provides tracking for charged particles within the
pseudorapidity region $|\eta| < 2.5$. At small radii, silicon pixel
layers and stereo pairs of silicon microstrip detectors provide
high-resolution position measurements. The pixel system consists of
four barrel layers, and three forward discs on either side of the
IP. The barrel pixel layers are positioned at radii of 33.3,
50.5, 88.5 and 122.5~mm. The silicon microstrip tracker (SCT)
comprises four double layers in the barrel and nine forward discs on
either side.  The radial position of the innermost (outermost) SCT
barrel layer is 299~mm (514~mm). The final component of the ID, the
transition-radiation tracker (TRT), is positioned at larger radii,
with coverage up to $|\eta|=2.0$.
 
The electromagnetic calorimeter is based on lead absorbers and liquid
argon and provides coverage for $|\eta|<3.2$.
Hadronic calorimetry uses steel absorbers and
scintillator tiles in the region $|\eta|<1.7$, and copper absorbers
with liquid argon in the endcaps ($1.5 < |\eta| < 3.2$). A
forward hadronic calorimeter using copper and tungsten absorbers with liquid
argon completes the calorimeter coverage up to $|\eta| = 4.9$.

The MS is the outermost ATLAS subdetector. It is designed to detect muons in the pseudorapidity region up to $|\eta| = 2.7$,
and to provide momentum measurements with a relative resolution better than 3\% over a wide $\pt$ range and up to 10\% at $\pt \approx 1$~\TeV.
The MS consists of one barrel ($|\eta| < 1.05$) and two endcap sections ($1.05 < |\eta| < 2.7$). A system of three large superconducting air-core toroidal magnets,
each with eight coils, provides a magnetic field with a bending integral of about $2.5$~Tm in the barrel and up to $6$~Tm in the endcaps. Resistive plate chambers (three doublet layers for $|\eta| < 1.05$) and thin gap chambers (one triplet layer followed by two doublets for $1.0<|\eta| < 2.4$) provide triggering capability to the detector as well as  ($\eta$, $\phi$) position measurements with typical spatial resolution of $5-10$~mm. A precise momentum measurement for muons with pseudorapidity up to $|\eta|=2.7$ is provided by three layers of monitored drift tube chambers (MDT). For $|\eta|>2$, the inner layer is instrumented with a quadruplet of cathode strip chambers (CSC) instead of MDTs. The single-hit resolution in the bending plane for the MDT and the CSC is about $80$~$\mu$m and $60$~$\mu$m, respectively. The muon chambers are aligned with a precision between $30$~$\mu$m  and $60$~$\mu$m.
 
A two-level trigger system is used to select
events~\cite{TRIG-2016-01}. The first-level trigger is implemented in
custom electronics and uses information from the muon trigger chambers
and the calorimeters. This is followed by a software-based high-level
trigger system, which runs reconstruction algorithms similar to those
used in offline reconstruction. Combined, the two levels reduce the
$40$~MHz bunch-crossing rate to approximately $1$~kHz of events saved
for further analysis.
 
 
\section{Prompt-trilepton signature}
\label{sec:pHNL}
 
The prompt-lepton search for HNLs is conducted in two channels: $W^\pm \rightarrow
\mu^\pm \mu^\pm e^\mp \nu_e$ (muon channel) and $W^\pm \rightarrow e^\pm e^\pm
\mu^\mp \nu_\mu$ (electron channel). It considers the case where the vertex
displacement is small enough that an ID track can be reconstructed
with standard ATLAS tracking algorithms. The standard reconstruction of tracks in the ID is optimised for charged particles that originate from the beam spot, the region where the proton beams intersect. This set-up restricts the detection for decay products of particles whose decay vertex is significantly displaced from the beam spot, especially for transverse displacements greater than approximately 20~mm~\cite{ATL-PHYS-PUB-2017-014}.
By requiring the final state to have three isolated leptons and no opposite-charge
same-flavour lepton pairs, background events from Drell--Yan pair production,
$W$+jets and $t \bar{t}$ could be rejected.
 
\subsection{Trigger and preselection (prompt signature)}
\label{pHNL:trigger}
 
Events are required to have a primary vertex defined as the reconstructed vertex with the largest sum of squared track transverse momenta.
For the muon channel, events are selected with a dimuon trigger with transverse momentum
$\pt$ thresholds of \SI{18} and \SI{8}{\GeV} for events recorded in 2015 and of
\SI{22} and \SI{8}{\GeV} for events recorded in 2016. For the electron channel, events are
selected with a logical OR between single-electron triggers with different transverse energy
thresholds and different electron identification criteria: either 24 {\GeV} (``medium''), 60 {\GeV} (``medium'') or \SI{120}{\GeV} (``loose'') for the 2015 dataset and either 26 {\GeV} (``tight''), 60 {\GeV} (``medium'') or \SI{140}{\GeV} (``loose'') for the 2016 dataset. The different identification criteria ``loose'', ``medium'' and  ``tight'' are defined in Ref.~\cite{electron_eff_paper}. A further trigger match requirement is imposed between the reconstructed leptons and the corresponding triggers.
 
Muon candidates are reconstructed from tracks in the MS,
matched with tracks found in the ID within $|\eta|<2.5$. Electron candidates are reconstructed from energy
deposits (clusters) in the electromagnetic calorimeter which are associated with
a reconstructed track in the ID, within the fiducial region
of $|\eta| <  2.47$, where $\eta$ is the pseudorapidity of the calorimeter
energy deposit associated with the electron candidate. Electron candidates within the
transition region between the barrel and endcap electromagnetic calorimeters,
$1.37 <|\eta|< 1.52$, are excluded. Muons are required to have a $\pt$ of at least \SI{4}{\GeV}, while the lowest $\pt$ threshold for electrons is
\SI{4.5}{\GeV} for 2016 data and \SI{7}{\GeV} for 2015 data. Both the muon and electron candidates are required to
satisfy ``loose'' sets of identification
criteria~\cite{ATLASmuons2016,electron_eff_paper} while for the electron channel the
electron that passed the single-electron trigger needs to satisfy ``tight'' identification criteria for
events in the dataset collected in 2016. The highest-$\pt$ (leading) lepton is required to pass
a transverse impact parameter significance requirement,
$d_0/\sigma(d_0)<3$
for muons and $<5$ for electrons, where the impact parameter $d_0$ is the transverse distance between the primary vertex and the point of closest approach of the lepton trajectory.
For other leptons, no such requirement is imposed to allow for a slight
displacement of leptons from HNLs. The leading lepton is also required to have
$|\Delta z_0\sin(\theta)|<0.5$~mm
while $<1$~mm is required for the remaining leptons, where $\Delta z_0$ is the distance along the beam direction between the primary vertex and the point of closest approach of the lepton trajectory. To
reduce the contribution from non-prompt leptons (e.g.\ from semileptonic $b$- or
$c$-hadron decays), photon conversions and hadrons, ``loose'' calorimeter and track isolation criteria as defined in Refs.~\cite{ATLASmuons2016,electron_eff_paper} are applied to lepton candidates, with a \SI{99}{\percent} efficiency. Scale factors are applied to simulated lepton distributions to take into account the small differences in reconstruction, identification, isolation, and trigger efficiencies between MC simulation and data. Energy/momentum scale and resolution corrections are also applied to leptons~\cite{electron_energyscale_paper}.
 
Jets are reconstructed from energy deposits in the calorimeter using the anti-$k_t$ algorithm~\cite{Cacciari:2008gp} with a radius parameter value of $R=0.4$. A multivariate technique (Jet Vertex Tagger)~\cite{ATLASjets2016} is applied in order to identify and select jets originating from the hard-scatter interaction, at a working point corresponding to a
\SI{92}{\percent} efficiency for identifying such jets and allowing an observed fake rate
of \SI{2}{\percent} from pile-up jets. Jets are also calibrated using energy- and $\eta$-dependent corrections~\cite{PERF-2016-04}. Only jets with a transverse momentum $\pt > 20$ GeV and within $|\eta| < 4.5$ are considered in the following. In order to identify jets containing $b$-hadrons (referred to as $b$-jets in the following), the MV2c10 algorithm is used, which combines impact parameter information with the explicit identification of secondary and tertiary vertices within the jet into a multivariate discriminant~\cite{ATL-PHYS-PUB-2016-012}. Operating points are defined by a single threshold in the range of the discriminant output values and are chosen to provide a specific $b$-jet efficiency in simulated $t\bar{t}$ events. The working point used in this analysis has an efficiency of \SI{77}{\percent} for $b$-jet tagging. It has rejection factors of 6, 22, and 134 against $c$-jets, hadronic decays of $\tau$-leptons, and jets from light quarks or gluons, respectively. A scale factor is applied to account for $b$-jet tagging efficiency differences between MC simulation and data~\cite{PERF-2016-05}.
 
The magnitude of the missing transverse momentum, $\etmiss$~\cite{ATLAS2018e}, is reconstructed from
the negative vector sum of transverse momenta of reconstructed and calibrated particles and jets. The reconstructed particles are electrons, photons, $\tau$-leptons and muons. Additionally, there is a second contribution calculated from ID tracks that are matched to the primary vertex and not associated with any of the selected objects (soft objects).
 
To avoid assigning a single detector response to more than one reconstructed
object, a sequential overlap-removal procedure is adopted. Jets are removed if
found to be within $\Delta R = 0.2$ of an electron or muon track, unless they satisfy the $b$-tagging requirements. In the electron channel, jets are not removed if their $\pt$ is at least
\SI{20}{\percent} higher than that of the electron. In the muon channel, a jet is not removed if it
has at least three tracks originating from the primary vertex. Electrons or muons
within a sliding-size cone around the remaining jets, defined as $\Delta R =$ min$ \{ 0.04 +
\SI{10}{\GeV} / \pt(\ell), 0.4 \}$, are rejected. Muons that can be matched to an energy deposit in the calorimeter (calorimeter-tagged muons) are removed if they share tracks with an electron. Electrons
are then removed if they share tracks with one of the remaining muons.

\subsection{Reconstruction and selection (prompt signature)}
\label{pHNL:selection}
 
All selected events are required to contain leptons which
must satisfy flavour and charge requirements, such that the event consists of
exactly $\mu^{\pm}\mu^{\pm}e^{\mp}$ in the muon channel and
$e^{\pm}e^{\pm}\mu^{\mp}$ for the electron channel. Furthermore, a requirement
is imposed on the three-lepton invariant mass constructed from the three final-state leptons:
\SI{40}{} $< m(\ell,\ell,\ell') <$ \SI{90}{\GeV}. Its distribution for signal events is centred  below the mass of the $W$ boson, as the neutrino escapes undetected.
 
If the mass of the HNL is smaller than half the mass of the $W$ boson, the
leading lepton will generally originate from the $W$ boson decay
and the other leptons from the $N$ decay. Accordingly, the dilepton invariant
mass $m(\ell,\ell')$ is determined as the invariant mass of the $e\mu$
combination which excludes the leading $e$ or $\mu$ in the electron and muon
channels respectively. Its distribution is centred  below the mass of the HNL.
 
In the muon channel, both the higher-$\pt$ and lower-$\pt$ (subleading) muons are required to have  $\pt > \SI{23}{\GeV}$ and $\pt > \SI{14}{\GeV}$,
respectively. In the electron channel, the leading and subleading electrons are required to have a transverse momentum
$\pt > \SI{27}{\GeV}$ and $\pt > \SI{10}{\GeV}$, respectively. Additionally, to reject the large number
of $Z$+jets background events that contain electrons with a misidentified charge, the invariant mass of the electrons must satisfy $m(e, e) <  \SI{78}{\GeV}$.
In both channels, the events must not have $b$-tagged jets and the $\etmiss$ value must be less than \SI{60}{\GeV}.
A summary of the signal region selection criteria is given in Table~\ref{table:pHNL signal region}.

\begin{table}[t]
\centering
\caption{Signal region selection criteria for the prompt trilepton analysis.}
\begin{tabularx}{0.7\linewidth}{XX}
\toprule
Muon channel                      &  Electron channel                        \\
\midrule
exactly $\mu^\pm \mu^\pm e^\mp$ signature  &  exactly $e^\pm e^\pm \mu^\mp$ signature          \\
\midrule
\multicolumn{2}{c} {$\pt(\mu) > \SI{4}{\GeV}$}  \\
\multicolumn{2}{c} {$\pt(e) >  \SI{7}{\GeV}$ (2015), \SI{4.5}{\GeV} (2016)}  \\
\midrule
leading muon $\pt > \SI{23}{\GeV}$         &  leading electron $\pt > \SI{27}{\GeV}$           \\
subleading muon $\pt > \SI{14}{\GeV}$     &  subleading electron $\pt > \SI{10}{\GeV}$       \\
&  $m(e, e) <  \SI{78}{\GeV}$                       \\
\midrule
\multicolumn{2}{c}{$\SI{40}{}  < m(\ell,\ell,\ell') <$ \SI{90}{\GeV}}                                  \\
\multicolumn{2}{c}{$b$-jet veto}                                                               \\
\multicolumn{2}{c}{$\etmiss < \SI{60}{\GeV}$}                                                  \\
\bottomrule
\end{tabularx}
\label{table:pHNL signal region}
\end{table}
 
\newenvironment{DIFnomarkup}{}{}
 
\begin{DIFnomarkup}
\begin{table}[b]
\centering
\caption{Prompt-trilepton signal efficiencies for muon and electron channels
after applying all selection criteria. The uncertainties are statistical only.}
\begin{tabularx}{\linewidth}{Xccccc}
\toprule
Channel & $m_N = \SI{5}{\GeV}$  & $m_N = \SI{10}{\GeV}$ & $m_N = \SI{20}{\GeV}$ & $m_N = \SI{30}{\GeV}$ & $m_N = \SI{50}{\GeV}$  \\
\textbf{} & $c\tau = \SI{1}{\mm}$ & $ c\tau = \SI{1}{\mm}$ & $ c\tau = \SI{0.1}{\mm}$ & $ c\tau = \SI{10}{\mu m}$ & $ c\tau = \SI{1}{\mu m}$  \\
\midrule
Muon  & (\SI{0.6}{}$\pm$0.1){\%} & (\SI{1.8}{}$\pm$0.2){\%} & (\SI{6.8}{}$\pm$0.4){\%} & (\SI{8.8}{}$\pm$0.5){\%}  & (\SI{9.9}{}$\pm$0.5){\%} \\
Electron  & (\SI{0.3}{}$\pm$0.1){\%} & (\SI{1.8}{}$\pm$0.2){\%} & (\SI{6.9}{}$\pm$0.4){\%} & (\SI{7.9}{}$\pm$0.5){\%}  & (\SI{5.1}{}$\pm$0.3){\%} \\
\bottomrule
\end{tabularx}
\label{table:pHNL efficiencies}
\end{table}
\end{DIFnomarkup}
 
\begin{figure}[tb]
\centering
\ifint
\includegraphics[width=0.6\linewidth]{fig_02a.png}
\else
\includegraphics[width=0.6\linewidth]{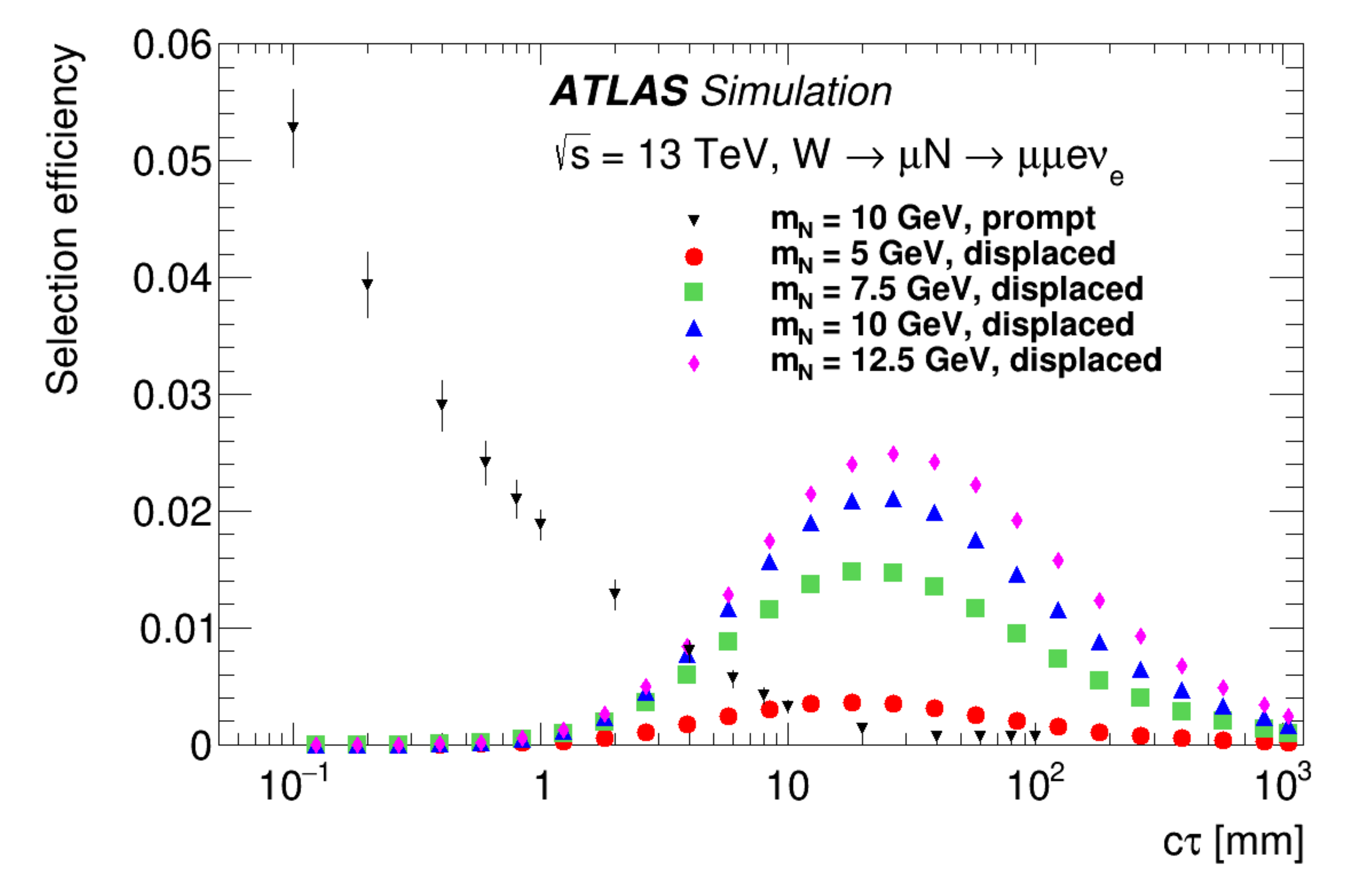}
\fi
\caption{HNL search event selection efficiency as a function of mean proper decay length evaluated from simulation. The efficiency for the prompt signature in the muon channel is shown for an HNL mass of 10 $\GeV$, while the efficiencies for the displaced signature are shown for four different HNL masses. Error bars represent MC statistical uncertainties.}
\label{fig:DVseleff}
\end{figure}
 
The impact of the final event selection on the efficiency for the signal samples
is shown in Table~\ref{table:pHNL efficiencies}. The selection efficiency is defined
as the fraction of generated events that were reconstructed and satisfied the selection criteria for the muon and electron channel ($pp \rightarrow W(\mu) N(\rightarrow \mu e \nu )$ or $pp \rightarrow W(e) N(\rightarrow e \mu \nu )$).
The black triangles in Figure~\ref{fig:DVseleff} show the
typical dependence of the final selection efficiency in the muon channel on the mean proper decay length of the HNL. The efficiency is constant for mean decay length up to about 0.1~mm, beyond which it decreases due to a sharp reduction in the efficiency of the standard tracking algorithm. This illustrates the complementarity with the displaced-vertex search described in Section~\ref{sec:dHNL}.

\subsection{Backgrounds and signal extraction (prompt signature)}
\label{pHNL:backgrounds}
 
The SM backgrounds that can lead to the same signature as the one from the
prompt heavy neutral lepton are a mixture of prompt real leptons and non-prompt leptons and leptons from pile-up.
These backgrounds can be split into two broad categories,
irreducible and reducible types. The irreducible background is composed of
exactly three leptons, where the only SM sources are diboson and triboson events as
well as $t\bar{t} V (V=W,Z)$. These backgrounds are negligible due to the
small cross section of these processes in this selection.
 
Reducible background events contain what is referred to as fake leptons.
In the case of electrons, these include contributions
from semileptonic decays of $b$- and $c$-hadrons, photon conversions and jets with large electromagnetic
energy (from hadronisation into $\pi^0$’s or from early showering in the calorimeter). In the case of
muons, they can originate from semileptonic decays of $b$- and $c$-hadrons, from charged hadron decays in the
tracking volume, or from punch-through particles emerging from high-energy
hadronic showers.
A large
fraction of such events originates from $W$+jets and multijets events, in the following referred to as multi-fake events. This background is determined from data.

Other subdominant contributions with two or more real leptons such as $Z$+jets, single-top-quark, diboson ($WW$, $WZ$ and $ZZ$) and triboson
events with hadronic decays are evaluated with simulation. The dominant $t\bar{t}$ background is derived from CRs in data, with shape extracted from MC simulation as described below.

To properly describe the backgrounds, a simultaneous binned maximum-likelihood fit is performed in three control regions (CR) and the signal region (SR). Normalisation factors are obtained for the most dominant MC background $\ttbar$ ($\muttbar$) and the multi-fake background ($\mudd$).
 
As described in Section \ref{pHNL:selection}, the SR is defined by selecting events with four criteria: 0) Same-charge and same-flavour (SCSF) lepton pairs, 1) \SI{40}{}   $< m (\ell, \ell, \ell') <$ \SI{90}{\GeV},
2) A $b$-jet veto and 3) $ \etmiss < \SI{60}{\GeV}$. The three CRs used are obtained by inverting one of the requirements of the SR definition (CR-1, CR-2 and CR-3). The maximum-likelihood fit uses the $\pt$ distribution of the third lepton for the three CRs, which provides good separation of the different background sources, while the $m(\ell,\ell')$ distribution is used for the SR, as it provides good separation of the background from the various signal models.
An overview of the criteria is given in Table~\ref{table:pHNL selection}.

\begin{table}[t]
\centering
\caption{Signal and control region selection criteria for the prompt HNL analysis and the corresponding distribution used in the binned maximum-likelihood fit in the SR (criterion 0) and the three CRs (criteria 1--3). In addition, estimation regions corresponding to the SR and the three CRs are defined by requiring all leptons to have the same charge. Criteria 0--3 are all used for the SR. Only one of them is inverted to define the corresponding CR.}
\begin{tabularx}{1\linewidth}{
>{\centering\arraybackslash}p{0.075\linewidth}
p{0.29\linewidth}
p{0.42\linewidth}
p{0.25\linewidth}
}
\toprule
Criterion & Signal region & Control region & Fit \\
&               &                & distribution \\
 
\midrule
0  & exactly one SCSF lepton pair  &    &  $m(\ell,\ell')$  \\
1  & \SI{40}{}   $< m(\ell, \ell, \ell') <$ \SI{90}{\GeV} &  $m(\ell, \ell, \ell') \leq \SI{40}{\GeV}\,\, ||\,\, m(\ell, \ell, \ell') \geq \SI{90}{\GeV}$  &   $\pt (\ell')$ \\
2  & $b$-jet veto                  &  at least one $b$-jet              &  $\pt (\ell')$   \\
3  & $ \etmiss < \SI{60}{\GeV}$    &  $\etmiss \geq \SI{60}{\GeV}$      &   $\pt (\ell')$   \\
\bottomrule
\end{tabularx}
\label{table:pHNL selection}
\end{table}

Three estimation regions (ER) are constructed the same way as the CRs except that all three leptons should have the same charge. In addition, a fourth estimation region is defined with only the requirement of all leptons having the same charge (and the rest of the SR requirements).
Additionally, isolation requirements on the leptons are loosened in all ERs to increase the number of events in data and also the number of fake leptons. The shape and initial normalisation of the multi-fake background are measured in these estimation regions as the difference between the data and the simulated backgrounds that contain two or more prompt real leptons. The third-lepton $\pt$ distribution is used for the three estimation regions corresponding to the three CRs, while the $m(\ell,\ell')$ distribution is used for the estimation region corresponding to the SR. The resulting statistical uncertainty dominates the multi-fake background uncertainties and is propagated as a bin-by-bin systematic uncertainty in the final maximum-likelihood fit.
 
A simultaneous binned maximum-likelihood fit is performed to data in order to obtain normalisation factors for the dominant $t\bar{t}$ background ($\muttbar$) and the derived multi-fake background ($\mudd$). A value
$\muttbar = 1.0$ would imply no normalisation change relative to the MC prediction, while $\mudd = 1.0$ would imply the same normalisation of the multi-fake background as was determined in the estimation regions.
Table~\ref{table:fit-parameters} shows the resulting normalisation factors from a fit under the background-only hypothesis using only the control regions \text{CR-1} to \text{CR-3}, as well as using these same three control
regions together with the signal region. The statistical uncertainty associated with multi-fakes in the control regions for the electron channel is large, giving enough freedom for the fit to estimate the normalisation factor. After adding the signal region the total statistical uncertainty is smaller because more data are added, constraining the normalisation factor. In the muon channel the statistical uncertainty is lower, reducing the impact of adding the signal region.

\begin{table}[b]
\centering
\caption{
Normalisation factors obtained for multi-fake $\mudd$ and
$t\bar{t}$ $\muttbar$ in different fit regions after the background-only
fit.
Shown are full
post-fit uncertainties of the normalisation factors taking into account all
statistical and systematic uncertainties.
}
\begin{tabularx}{0.6\linewidth}{cccc}
\toprule
Channel& Fit configuration & $\mudd$ & $\muttbar$ \\
\midrule
\multirow{ 2}{*}{Muon} & only \text{CR}s  & \SI{0.97 \pm 0.38}{} & \SI{0.90 \pm 0.14}{} \\
& \text{CR}s + \text{SR}  & \SI{1.48 \pm 0.34}{} & \SI{0.88 \pm 0.13}{} \\
\midrule
\multirow{ 2}{*}{Electron} & only \text{CR}s & \SI{0.42 \pm 0.92}{} & \SI{1.02 \pm 0.16}{}  \\
& \text{CR}s + \text{SR} & \SI{0.81 \pm 0.28}{} & \SI{0.99 \pm 0.15}{}  \\
\bottomrule
\end{tabularx}
\label{table:fit-parameters}
\end{table}
 
The total uncertainty in the expected signal yield is presented for the different signal points
in Table~\ref{table:pHNL uncertainty} using the post-fit systematic uncertainties. The relative statistical uncertainty arising from the limited number of events in the MC samples and in the CRs for the multi-fake background estimation is shown in parentheses. Systematic
uncertainties do not have a large impact on the signal yield except for the lowest masses. The most prominent
systematic uncertainties are related to the variation of the $\pt$ resolution correction of muon tracks in the
MS and ID, the variation of the energy scale calibration of
electrons, and calibration variations of the jet energy resolution and scale.

\begin{table}[tb]
\centering
\caption{Prompt-trilepton relative uncertainty of signal yields for muon and electron channels
after applying the selection criteria. Uncertainties correspond to those obtained after the fit. Uncertainties associated to the limited number of events in the MC samples and in the CRs for the multi-fake background estimation are shown in parentheses.}
\begin{tabularx}{\linewidth}{Xccccc}
\toprule
Channel & $m_N = \SI{5}{\GeV}$  & $m_N = \SI{10}{\GeV}$ & $m_N = \SI{20}{\GeV}$ & $m_N = \SI{30}{\GeV}$ & $m_N = \SI{50}{\GeV}$  \\
\textbf{} & $c\tau = \SI{1}{\mm}$ & $ c\tau = \SI{1}{\mm}$ & $ c\tau = \SI{0.1}{\mm}$ & $ c\tau = \SI{10}{\mu m}$ & $ c\tau = \SI{1}{\mu m}$  \\
\midrule
Muon  & \SI{19}{\percent} (\SI{15}{\percent}) & \SI{8.1}{\percent} (\SI{6.7}{\percent}) & \SI{4.2}{\percent} (\SI{4.1}{\percent})  & \SI{4.3}{\percent} (\SI{3.9}{\percent}) & \SI{3.5}{\percent} (\SI{3.3}{\percent})  \\
Electron  & \SI{14}{\percent} (\SI{14}{\percent}) & \SI{7.8}{\percent} (\SI{7.0}{\percent}) & \SI{4.8}{\percent} (\SI{4.1}{\percent}) & \SI{4.6}{\percent} (\SI{3.6}{\percent})  & \SI{4.7}{\percent} (\SI{3.6}{\percent}) \\
\bottomrule
\end{tabularx}
\label{table:pHNL uncertainty}
\end{table}

For all simulated backgrounds, systematic uncertainties related to the reconstruction and
identification of leptons and jets as well as the missing transverse momentum are
applied. Uncertainties associated with charged leptons arise from the trigger selection, the object reconstruction, the identification, and the isolation criteria, as well as the lepton momentum scale and resolution~\cite{ATLASmuons2016, electron_eff_paper}. Uncertainties associated with jets arise from the jet reconstruction and identification efficiencies related to the jet energy scale (JES) and jet energy resolution, and from the Jet Vertex Tagger efficiency~\cite{PERF-2016-04}. The JES-related uncertainties, contain 23 components which are treated as statistically independent and uncorrelated. Some of these components are related to jet flavour, pile-up corrections, $\eta$ dependence and high-$\pt$ jets. The uncertainties associated with $\etmiss$ are propagated from the uncertainties in the reconstruction of leptons and jets since they are used for the $\etmiss$ calculation. Uncertainties due to soft objects are also considered~\cite{ATLAS2018e}.
Additional cross-section uncertainties of the various backgrounds estimated in MC simulation are considered. The systematic
uncertainties of the backgrounds are smaller than the statistical
uncertainty in the phase space selected by this prompt HNL
analysis.
The most prominent systematic uncertainties are related to the identification of the analysis
objects, namely (relative uncertainty in parentheses)
calibration variations of the jet energy resolution and scale
(\SI{6}{\percent}), jet pseudorapidity inter-calibration (\SI{5}{\percent}) and
efficiency scale factor of $b$-jet tagging (\SI{2}{\percent})~\cite{PERF-2016-05}. Other systematic uncertainties
such as the variation of the $\pt$ resolution correction of muon tracks in the
MS and ID and the variation of the energy scale for the calibration of
electrons~\cite{electron_energyscale_paper} account for systematic effects
smaller than $\sim$1\%.
A \SI{50}{\percent} systematic uncertainty is also applied to
the yield of $Z$+jets events, accounting for possible mismodelling of non-prompt
leptons in the simulation.
 
Systematic uncertainties are parameterised by nuisance parameters with Gaussian constraints in the likelihood fit. These nuisance parameters are shared between all samples, while statistical uncertainties are modelled using Gamma distributions. The statistical uncertainty of the total MC-based background is treated with a single nuisance parameter for each bin in the likelihood~\cite{Cranmer:1456844,Baak:2014wma}.

\begin{figure}[tb]
\centering
\ifint
\includegraphics[width=0.329\linewidth]{fig_03a.png}
\includegraphics[width=0.329\linewidth]{fig_03b.png}
\includegraphics[width=0.329\linewidth]{fig_03c.png}
\else
\includegraphics[width=0.329\linewidth]{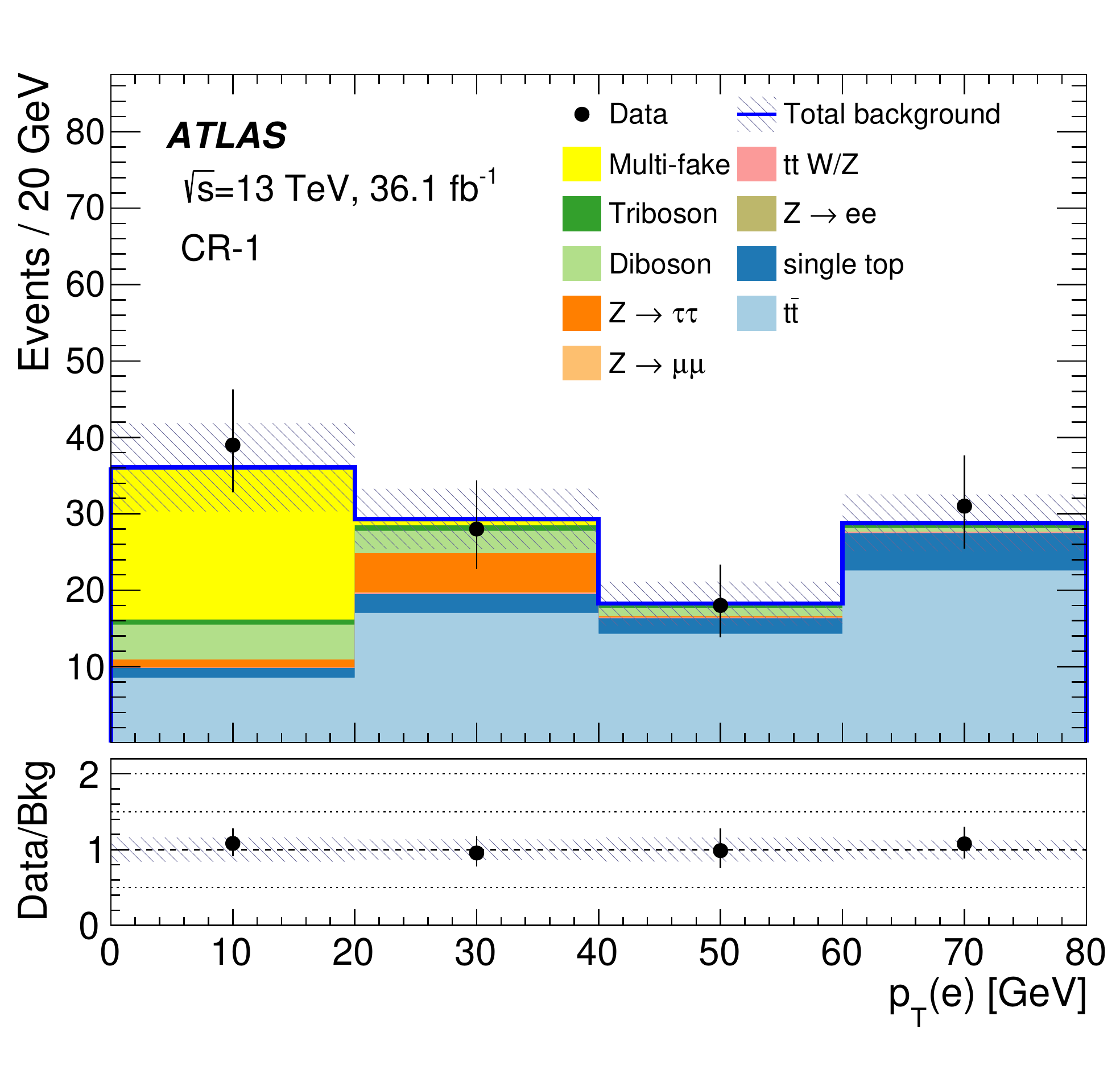}
\includegraphics[width=0.329\linewidth]{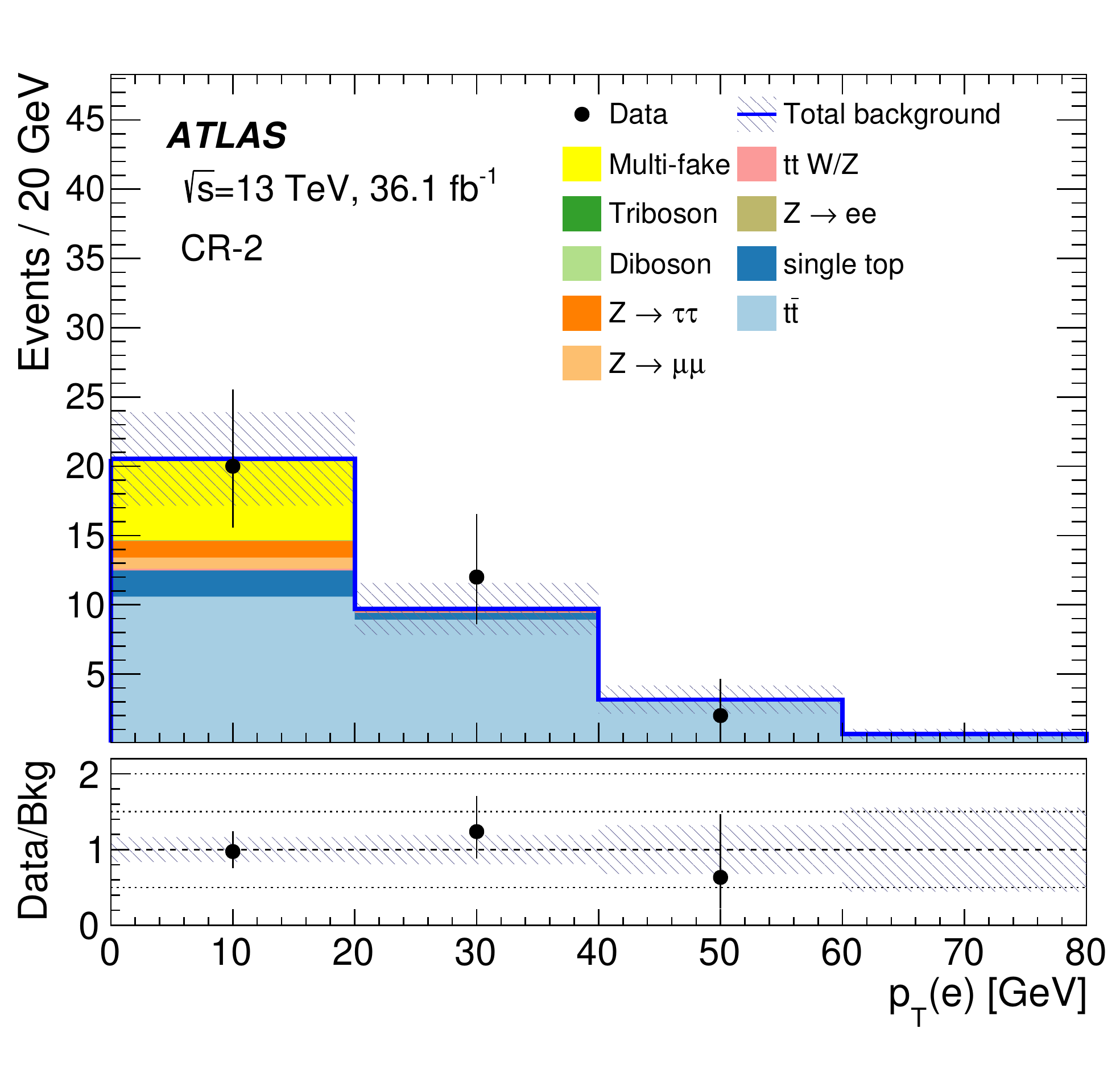}
\includegraphics[width=0.329\linewidth]{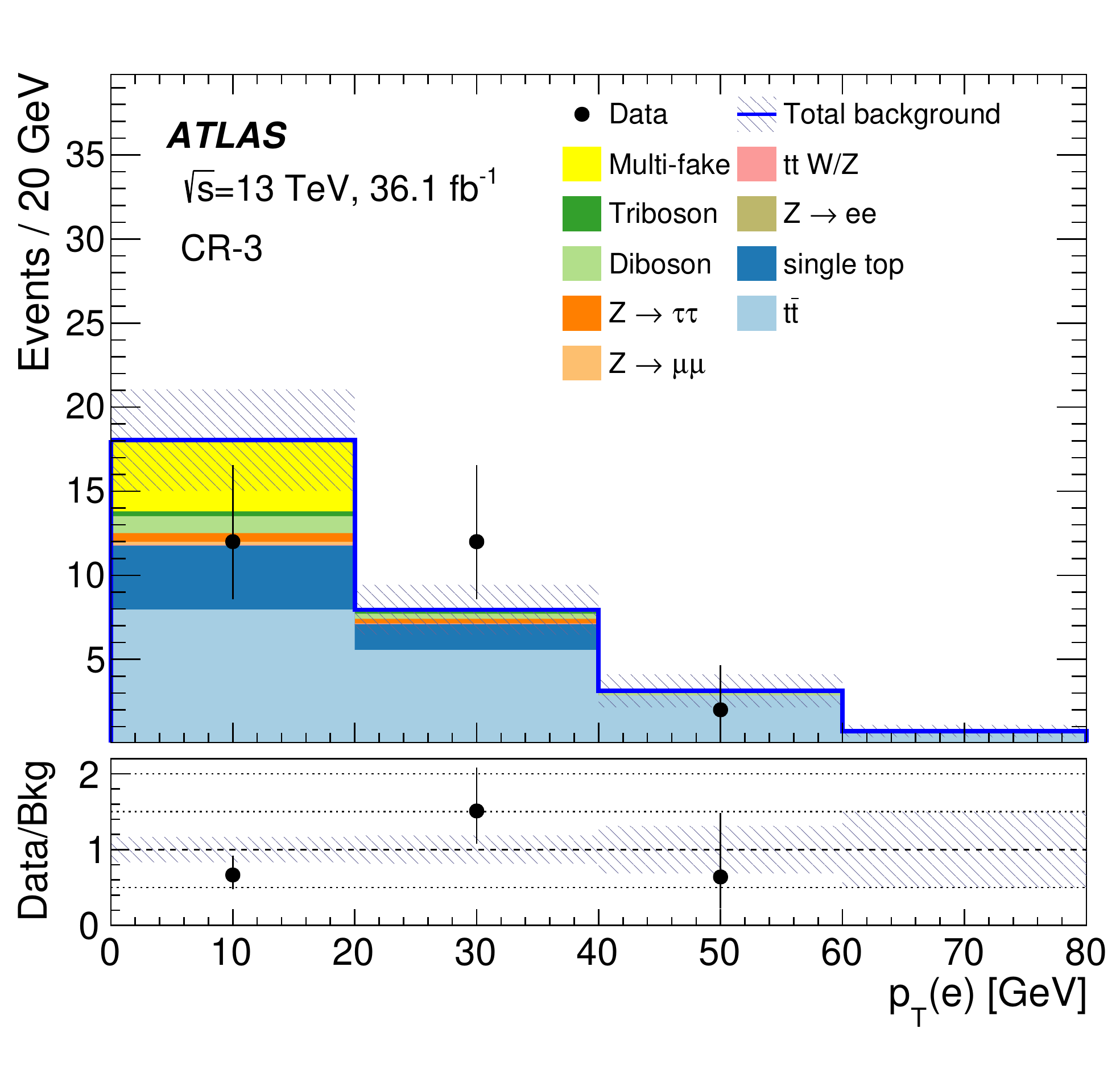}
\fi
\caption{Third-lepton $\pt$ distributions in the control regions CR-1 (left), CR-2 (middle) and CR-3 (right) for the prompt HNL analysis in the muon channel, showing post-fit background-only hypothesis including all the uncertainties and normalised in both \text{CR}s+\text{SR}. The total uncertainty in the background is shown as dashed regions. }
\label{fig:post-fit:mu}
\end{figure}

\begin{figure}[tb]
\centering
\ifint
\includegraphics[width=0.329\textwidth]{fig_04a.png}
\includegraphics[width=0.329\textwidth]{fig_04b.png}
\includegraphics[width=0.329\textwidth]{fig_04c.png}
\else
\includegraphics[width=0.329\textwidth]{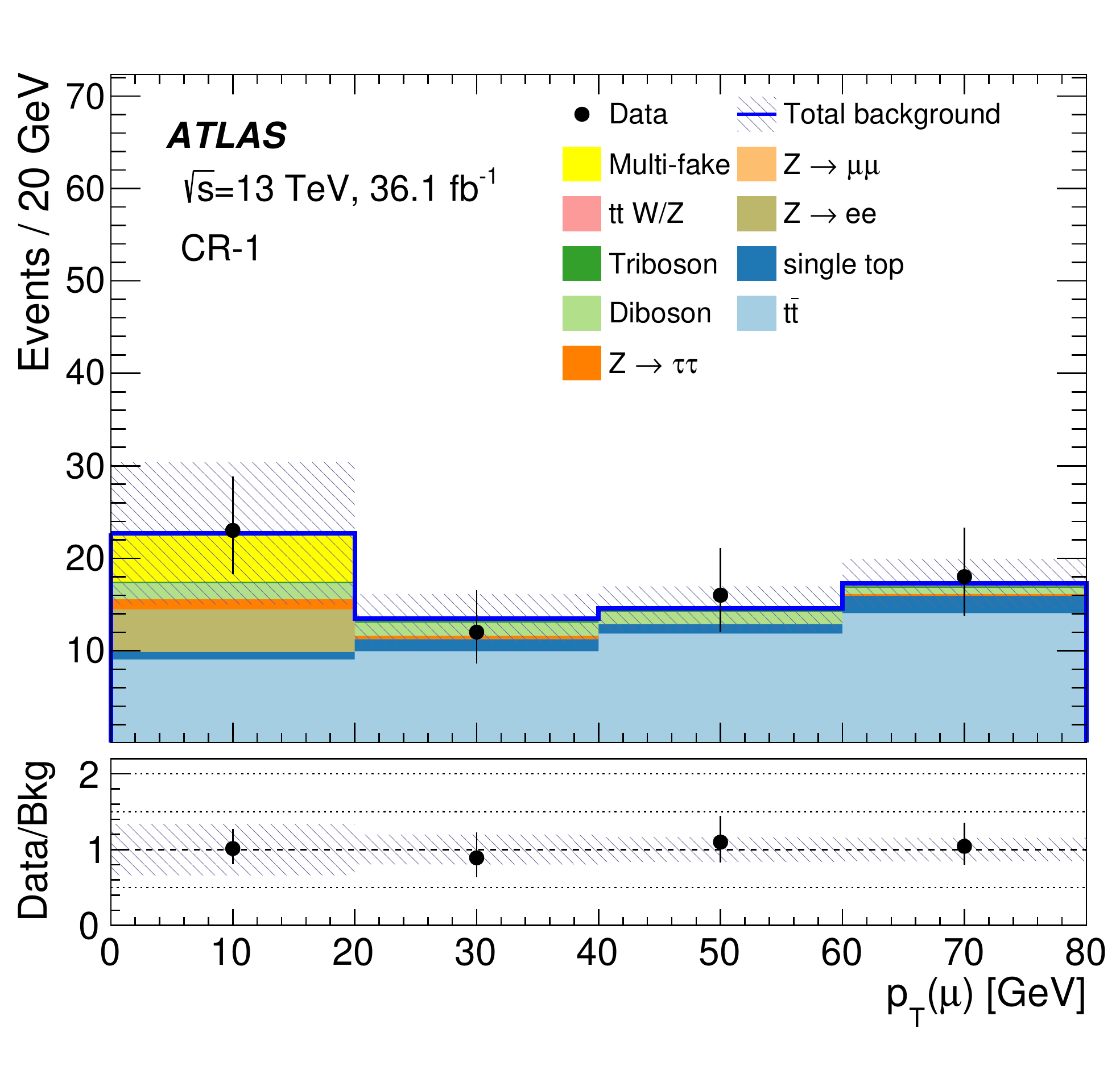}
\includegraphics[width=0.329\textwidth]{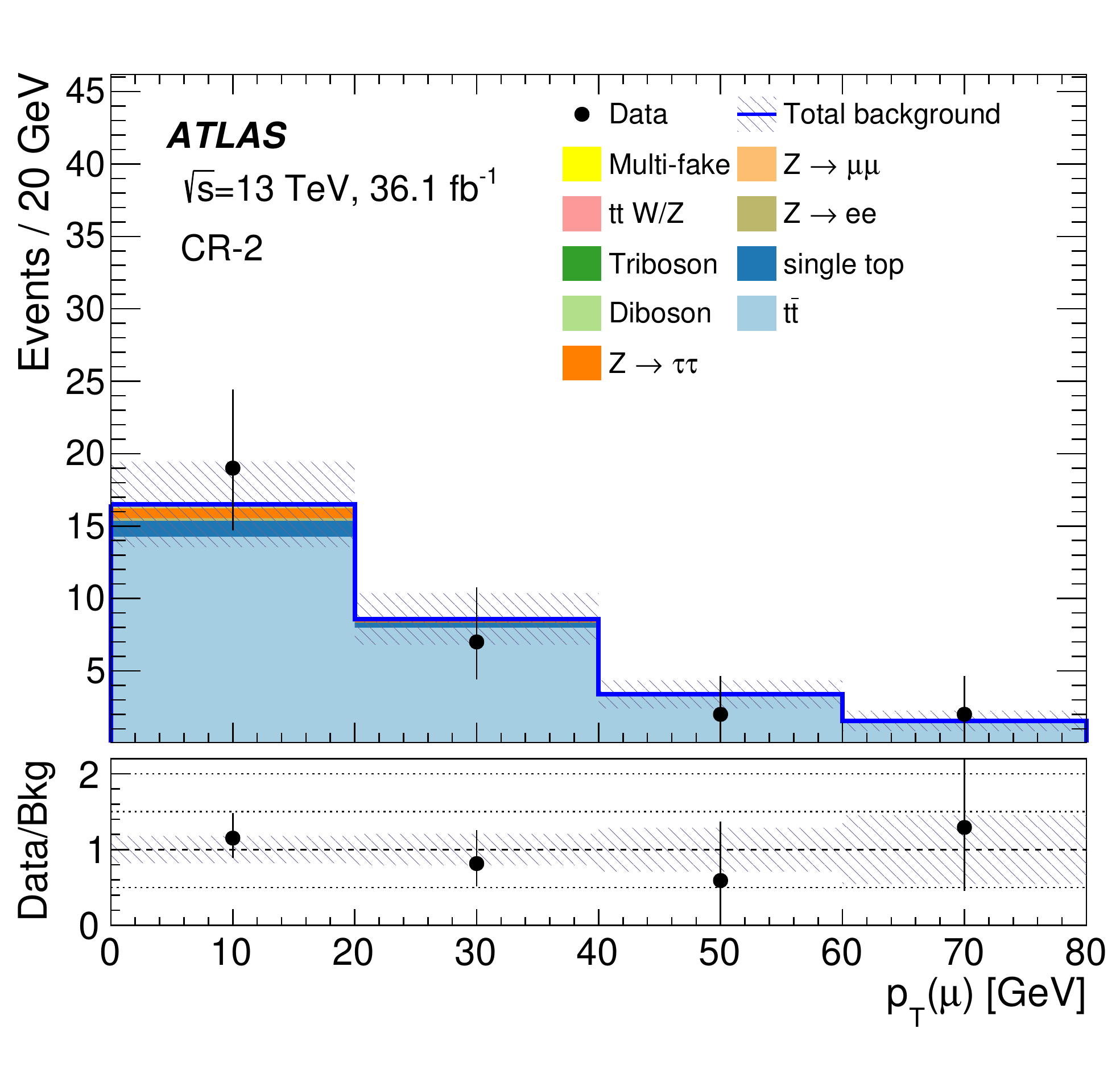}
\includegraphics[width=0.329\textwidth]{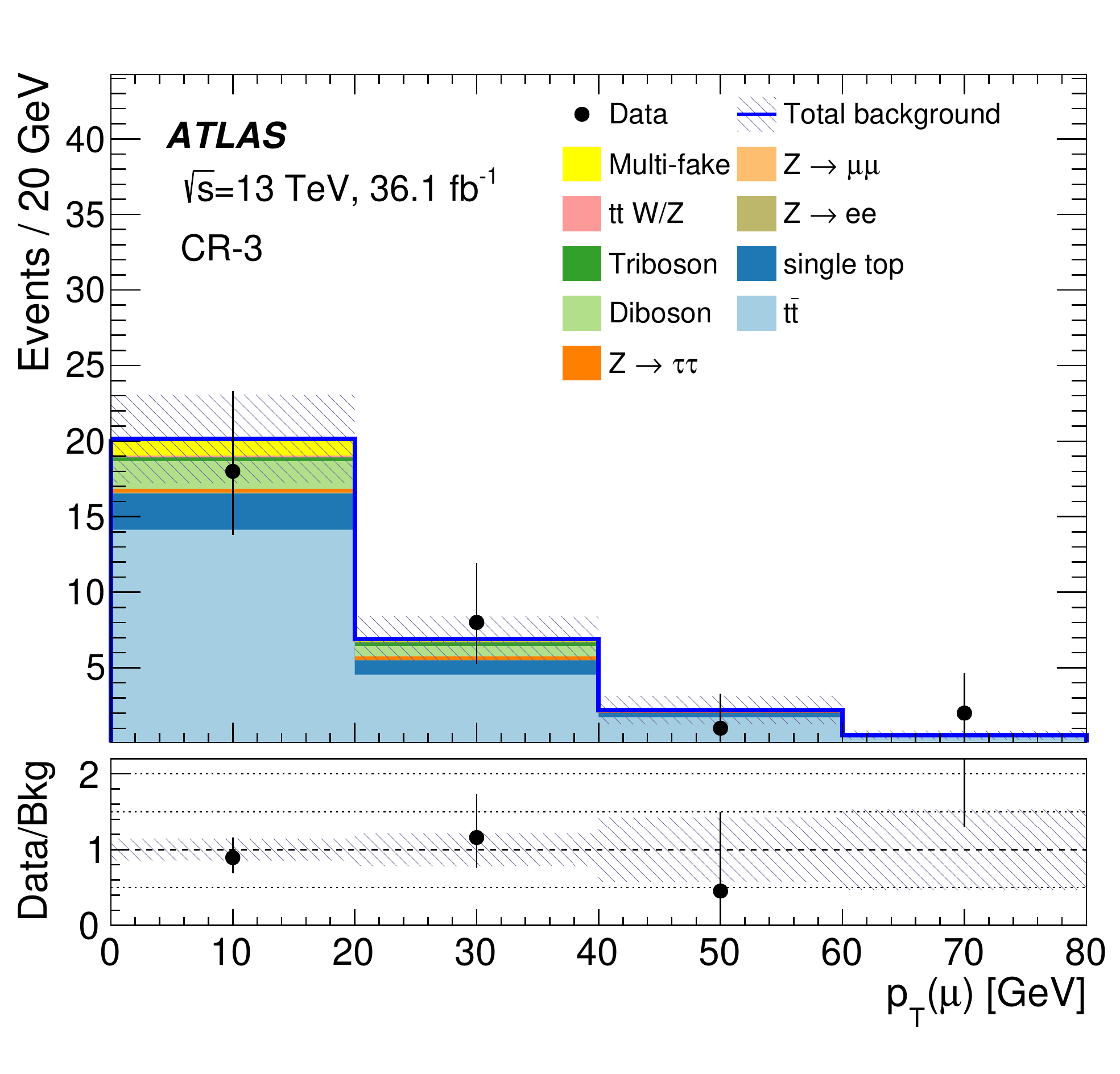}
\fi
\caption{Third-lepton $\pt$ distributions in the control regions CR-1 (left), CR-2 (middle) and CR-3 (right) for the prompt HNL analysis in the electron channel, showing post-fit background-only hypothesis including all the uncertainties and normalised in both \text{CR}s+\text{SR}. The total uncertainty in the background is shown as dashed regions. }
\label{fig:post-fit:el}
\end{figure}

\begin{figure}[tb]
\centering
\ifint
\includegraphics[width=0.45\textwidth]{fig_05a.png}
\hspace{0.5cm}
\includegraphics[width=0.45\textwidth]{fig_05b.png}
\else
\includegraphics[width=0.45\textwidth]{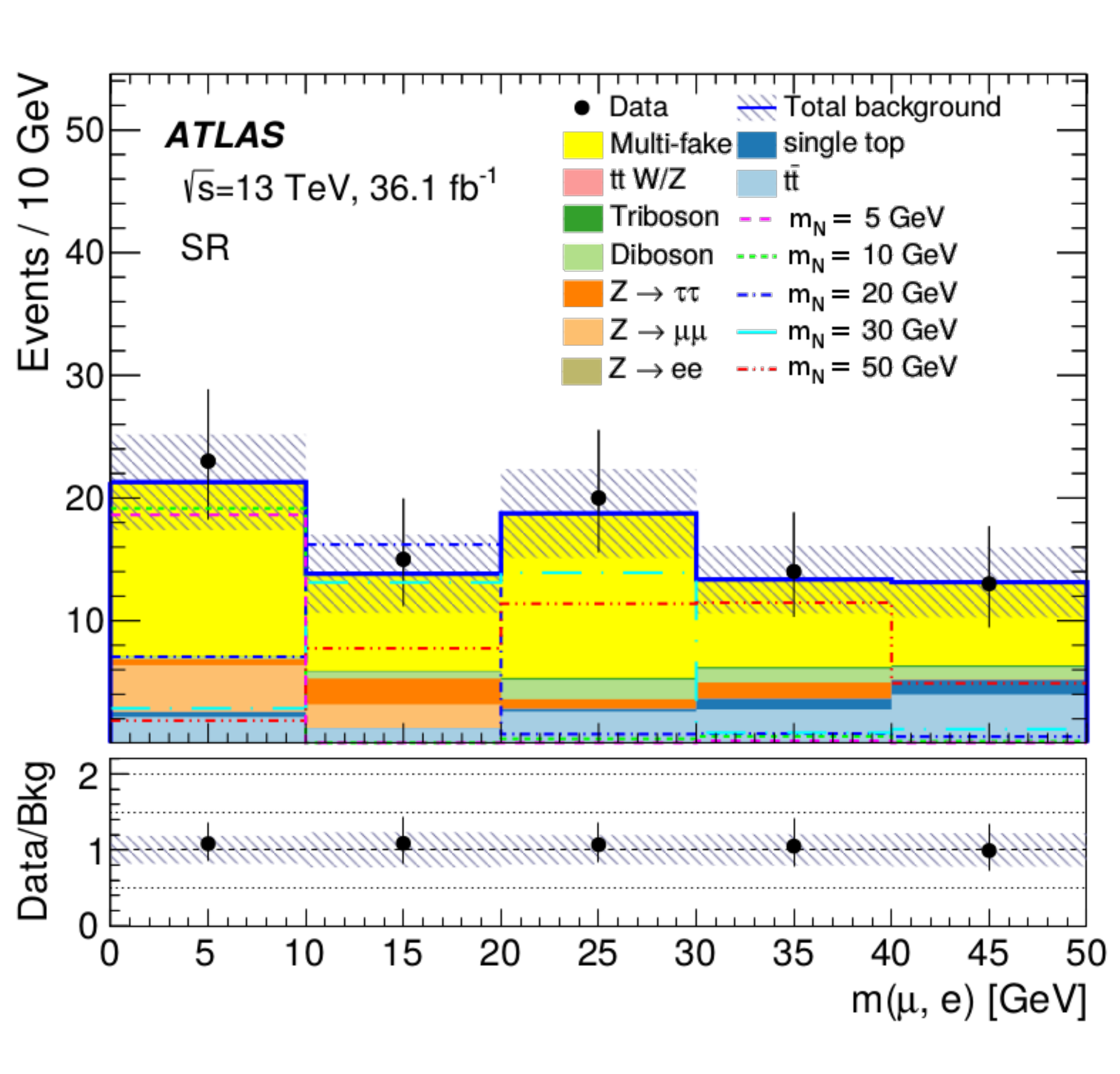}
\hspace{0.5cm}
\includegraphics[width=0.45\textwidth]{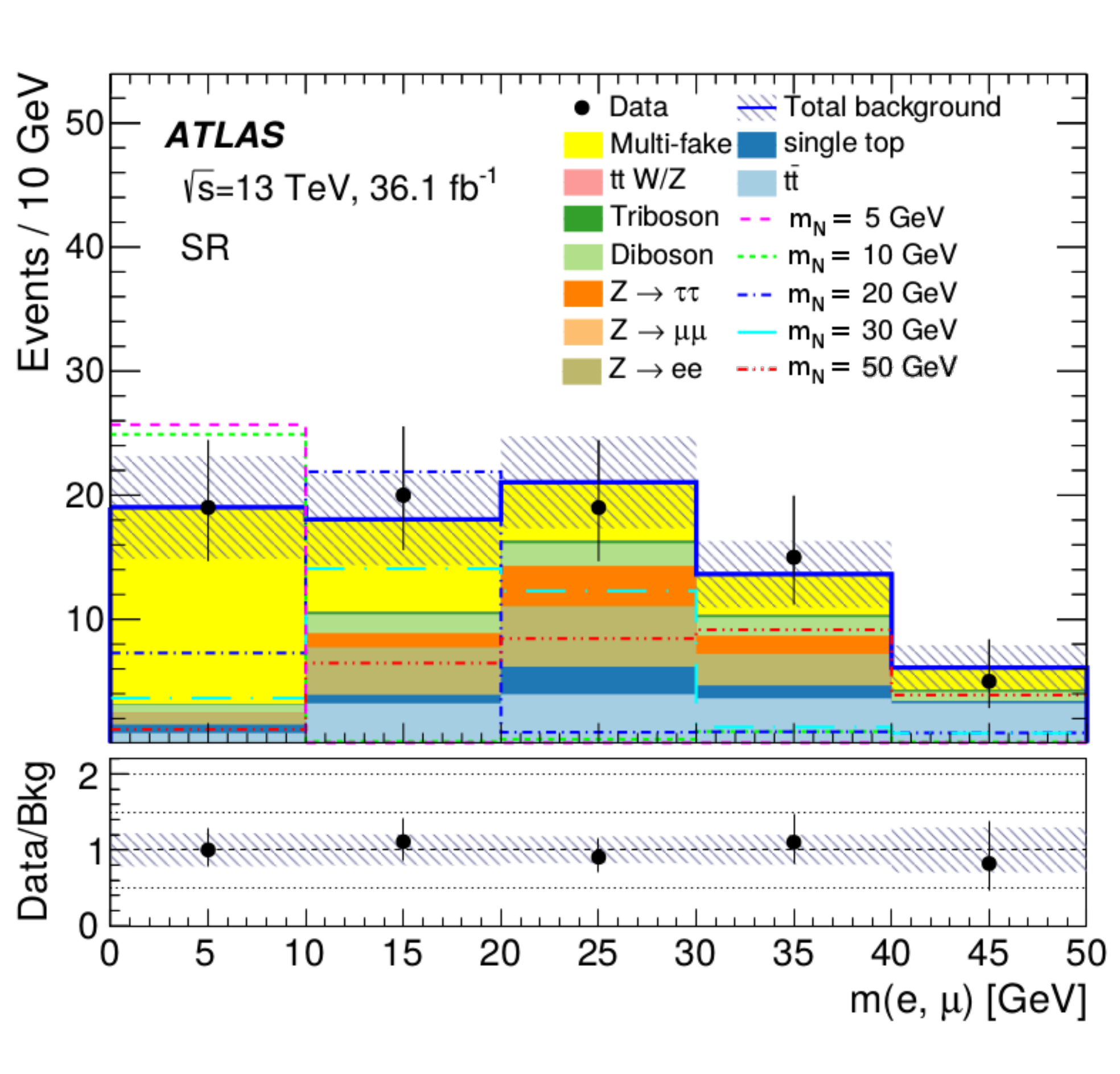}
\fi
\caption{Data-to-prediction comparison of dilepton invariant mass distributions in the signal regions for the prompt signature in the muon (left) and electron (right) channels. The prediction including its uncertainty is post-fit. The total uncertainty in the background is shown as dashed regions. Signal yields, overlaid as lines, correspond to observed 95\% confidence-level exclusion limits for different mass scenarios: for the muon channel, the signals have ($m_N$, $c\tau$, $|U|^2$) values of (5 GeV, 4.5 mm, 7.1 $\times 10^ {-5}$), (10 GeV,  2.7 $\times 10^ {-1}$ mm, 3.2 $\times 10^ {-5}$), (20 GeV, 2.1 $\times 10^ {-2}$ mm, 1.1 $\times 10^ {-5}$), (30 GeV, 2.2 $\times 10^ {-3}$ mm, 1.3 $\times 10^ {-5}$) and (50 GeV, 8.2 $\times 10^ {-5}$ mm, 2.5 $\times 10^ {-5}$); for the electron channel, the signals have ($m_N$, $c\tau$, $|U|^2$) values of (5 GeV, 1.6 mm, 1.9 $\times 10^ {-4}$), (10 GeV, 2.1 $\times 10^ {-1}$ mm, 4.1 $\times 10^ {-5}$), (20 GeV, 1.6 $\times 10^ {-2}$ mm, 1.4 $\times 10^ {-5}$), (30 GeV, 1.9 $\times 10^ {-3}$ mm, 1.5 $\times 10^ {-5}$) and (50 GeV, 5.4 $\times 10^ {-5}$ mm, 3.8 $\times 10^ {-5}$).
}
\label{fig:signal-overlay}
\end{figure}

The third-lepton $\pt$ distributions for each control region using the post-fit normalisation factors as measured in
control and signal regions combined are shown in Figures~\ref{fig:post-fit:mu}
and \ref{fig:post-fit:el} for the muon and electron channels, respectively. The number of multi-fake events in the electron channel is about a third of the ones in the muon channel. When performing the CR-only fit, this is reduced further by a factor of about two, yielding a large uncertainty in the electron channel $\mudd$ CR-only result.
The invariant mass distributions of the second and third leptons in the signal region for the signal-plus-background post-fit are shown in Figure~\ref{fig:signal-overlay}. This variable provides a good signal and background separation as most signal mass points concentrate in the first three bins, also allowing for distinction between different signal mass values.

The difference between data and the multi-fake estimate in the lowest bin of \text{CR-3} (see Figure \ref{fig:post-fit:mu}, right) is responsible for a change of $\mudd$ by 1.3$\sigma$ when including the \text{SR} in the fit for the muon channel (see Table~\ref{table:fit-parameters}). This was found to be anti-correlated with the statistical uncertainty parameterisations of MC and multi-fake backgrounds in the lowest bin of \text{CR-3}, as a consequence of a statistical fluctuation of data in the low-$\pt$ spectrum of this region. Further checks were conducted, including a $p$-value test in the signal region using normalisation factors as measured in the \text{CR}s. The $p$-values are determined without taking into
account the fit parameters associated with statistical uncertainty in the \text{SR} and found to be \SI{7.2}{\percent}
for the muon channel and \SI{13.2}{\percent} for the electron channel. As this
lies above the usual rejection level of \SI{5}{\percent} the test is considered
satisfactory. Upon using the background-only hypothesis normalised in both
\text{CR}s and \text{SR} very good compatibility was found.
 
For this search, two models are compared with the data: the background-only model describing the
SM processes and a signal-plus-background model in which the signal under
investigation is added to the SM background. First a discovery hypothesis
test is performed in which a rejection of the background-only model given the
observed data is examined. Afterwards the signal strength in the
signal-plus-background model is fixed to different values and the
CL$_{\mathrm{S}}$~\cite{Read:2002hq} method
is used to exclude various signal strengths. 
Exclusion limits on the signal-strength fit parameter
translate directly into limits on $|U|^2$. A signal
with a given mixing angle is considered excluded once the $p$-value for the
CL$_{\mathrm{S}} = $CL$_{\mathrm{s+b}}/$CL$_{\mathrm{b}}$ value is below 0.05 and therefore corresponds to a
\SI{95}{\percent} confidence-level exclusion limit. An overview of different signal shapes in
the \text{SR} is shown in Figure~\ref{fig:signal-overlay} for which the
signal yield corresponds to a \SI{95}{\percent} confidence-level exclusion limit for each mass
point. The exclusion limits in $|U_{\mu}|^2$ and $|U_{e}|^2$ for the different signal hypotheses are shown in Figure~\ref{fig:limits} together with those obtained from the displaced HNL search. The expected limits are determined using the results of a CR-only fit as input to the limit extraction procedure.
The offset by almost $1\sigma$ from the expected limit for the muon channel
shown in Figure~\ref{fig:limits} originates from the use of central values for the
statistical uncertainty Gamma distributions in the \text{SR} for the Asimov
dataset. In Figure~\ref{fig:signal-overlay} scaled (post-fit) Gamma nuisance parameters are
used for the statistical uncertainty parameters, which largely dominate over systematic uncertainties. For the electron channel the
post-fit statistical uncertainty parameters deviate less from their
central value as $\mudd$ has more freedom in the \text{CR}s, and consequently a
general offset is not observed. The deviation for the muon channel is a
consequence of the aforementioned $\mudd$ constraint in the lowest bin of
\text{CR-3}.
 
\FloatBarrier
 
\section{Displaced-vertex signature}
\label{sec:dHNL}
 
For $m_N\lesssim 20$~GeV, as the HNL lifetime gets longer for lower masses and coupling strengths, the searches relying on standard prompt objects, as defined above, become highly inefficient. The signature of a displaced vertex (DV) is needed to explore these complementary regions of the parameter space. Another advantage of this approach is that the requirement of a DV detached from the primary $pp$ interaction by 4~mm or more in the transverse plane eliminates the vast majority of SM backgrounds. In addition to probing lower masses and coupling strengths, this allows the requirement of same-charge same-flavour leptons to be removed and thus the search to be performed without relying on LNV.
 
Searches for long-lived neutral particles using DV signatures conducted so far by the ATLAS~\cite {ATLAS2012b,ATLAS2013c,ATLAS2014a,ATLAS2015b,ATLAS2017a,ATLAS2018d,ATLAS2018c,ATLAS2018b} and CMS~\cite{CMS-EXO-12-037,CMS-EXO-12-038,CMS2014b,CMS2018e,CMS2018d,CMS2018c,CMS2018b} Collaborations considered the new particles to be high-$\pt$ decay products of other massive particles, with relatively large branching ratios. None of these provided any relevant sensitivity to HNLs due to high-$\pt$ requirements or to the requirement that two displaced vertices must be reconstructed in the same event. It was noted~\cite{Helo2014,Izaguirre2015,Mermod2017} that HNLs from $W$ boson decays produce a distinct signature which had not yet been probed at the LHC: a prompt lepton from the $W$ boson and a DV formed with tracks of relatively low $\pt$, among which there is at least one belonging to a lepton.
 
\subsection{Trigger and preselection (displaced signature)}
\label{dHNL:trigger}
 
The displaced signature explored in this search comprises a prompt and isolated muon accompanied by a DV formed by either two muons or a muon and an electron. A prompt lepton from the $W$ boson decay is essential for triggering the event. The single-muon trigger with a $\pt$ threshold of 26~GeV is used.
 
Displaced vertices in the ID can be reconstructed at radial distances up to the first SCT layer at $\sim$300~mm due to the application of a large-radius tracking (LRT) algorithm optimised for tracks with large unsigned transverse impact parameters relative to the primary vertex ($d_0$)~\cite{ATLAS2017b}. Large-radius tracking is computationally intensive and is therefore not performed on the entirety of the dataset but rather on a subset of events preselected according to a specific set of criteria which must rely on the standard track-reconstruction algorithm. In this analysis, this preselection requires the presence of at least two muons with $|\eta|<2.5$ and is applied to the 2016 dataset, corresponding to an integrated luminosity of 32.9~fb$^{-1}$. One muon, meant to originate from an on-shell $W$ boson decay, is required to have an ID track matched to an MS track segment, have $\pt > 28$~GeV, and satisfy ``loose'' isolation and identification criteria as defined in Ref.~\cite{ATLASmuons2016}. Another muon, which targets a displaced muon from an HNL decay, is required to satisfy a ``loose'' isolation criterion and have $\pt > 5$~GeV; it must have an MS track which either has no matched track in the ID or, if it has a matched track, must have either $d_0>0.1$~mm, or $d_0<0.1$~mm and $\chi^2$/dof~$>5$ for the MS--ID track matching, as defined in Ref.~\cite{ATLASmuons2016}. The $\chi^2$/dof selection is added to recover tracks which have relatively low $d_0$ values but still correspond to a displaced muon. For instance, it could be that, during standard reconstruction, the MS track from a muon is incorrectly matched to an unrelated ID track.
 
\subsection{Reconstruction and selection (displaced signature)}
\label{dHNL:selection}

Large-radius tracking is performed on the dataset which satisfies the preselection, producing an extended collection of tracks. Using these new tracks, muon and electron candidates are reconstructed as described in
Section~\ref{pHNL:trigger}.
Displaced-vertex candidates are also formed using this extended track collection. The following additional requirements are imposed on the tracks to be considered as seeds of the vertex reconstruction algorithm: they must have $\pt > 1$~GeV, $d_0>2$~mm, at least two SCT hits, and at least one TRT hit or two pixel hits. The vertex reconstruction algorithm described in Ref.~\cite{ATLAS2017a} is used, based on the \textit{incompatibility graph method} and iteratively merging all two-track vertices which are found within 1~mm distance from each other. For an HNL decay length of the order of a cm, the DV reconstruction efficiency (including the track reconstruction efficiency) is about 20\%, as studied in Ref.~\cite{ATLAS2019b}.
 
The event selection requires the presence of at least one DV which satisfies the properties described below. The DV must be within the fiducial volume defined as $4 < r_{\textrm{DV}} < 300$~mm, where $r_{\textrm{DV}}$ is the distance to the beam axis. Since TRT hits do not have a well defined $z$ coordinate, the upper bound on $r_{\textrm{DV}}$ ensures that tracks have a sufficient number of SCT hits for high-quality track and DV reconstruction. Then the DV is required to be formed by exactly two tracks with opposite charges in order to be identified as decay products of a neutral particle. A cosmic-ray veto, $\sqrt{(\sum\eta)^2 + (\pi-\Delta\phi)^2} > 0.04$ (in which the sum $\sum\eta$ and difference $\Delta\phi$ refer to the two tracks forming the DV), is applied to eliminate high-mass vertices from a single cosmic-ray muon which is reconstructed as two back-to-back muons. The DV must be formed by at least one tight muon and an additional tight lepton (either muon or electron), with a ``tight'' identification identical to the standard one~\cite{ATLASmuons2016,electron_eff_paper} except that it does not impose a minimum number of pixel hits such as to be efficient for DVs originating beyond the first pixel layer. Given a reconstructed track originating from the HNL decay, tight-lepton reconstruction efficiencies are around 70\% and 50\% for muons and electrons, respectively. Finally, a requirement is applied on the DV invariant mass $m_{\textrm{DV}}$ as defined by $m^2_{\textrm{DV}}=(\sum E_i)^2 - (\sum \vec{p_i})^2$, where the sum runs over the two tracks forming the vertex. The requirement is chosen to be $m_{\textrm{DV}}>4$~GeV as a compromise between keeping good signal efficiencies for HNL masses of 5~GeV and above, and rejecting low-mass backgrounds from material interactions and decays of metastable SM particles.

The overall signal efficiency, defined as the fraction of generated signal events that were reconstructed and satisfied the selection criteria, depends on the HNL mass and lifetime and is typically (1--2)\% in the regions probed by the displaced HNL search. Signal losses are largely dominated by inefficiencies for displaced tracks and DV reconstruction.
For a given HNL mass, the efficiency for any value of the mean proper decay length $c\tau$ is  obtained from  the simulated samples generated with $c\tau=$ 1, 10, 100~mm. To evaluate the efficiency for a  given  value  of $c\tau$, each reconstructed event is weighted so that the generated event sample is distributed according to $\exp(-t_{\textrm true}/\tau)$, where $t_{\textrm true}$ is the true  proper  decay time.
The resulting efficiencies are shown in Figure~\ref{fig:DVseleff}. The efficiencies evaluated with this method agree with those from the fully simulated MC samples with $c\tau$ of 1, 10 and 100~mm within statistical uncertainties. Efficiencies increase with increasing $m_N$ due to the requirement on the reconstructed vertex mass $m_{\textrm{DV}}>4$~GeV and due to the fact that the decay leptons have larger momenta. The search is not sensitive to values of $c\tau$ lower than  0.1~mm or higher than 1000~mm due to the requirement of a DV in the fiducial volume
$4 < r_{\textrm{DV}} < 300$~mm.
 
Relevant systematic uncertainties that can affect the signal efficiencies (with relative magnitude in between parentheses) include uncertainties in the efficiencies for the prompt-lepton reconstruction and identification (1\%), displaced track and DV reconstruction (15\%), displaced-lepton identification (5\%), as well as uncertainties in the modelling of lepton kinematic distributions and individual decay branching ratios (10\%), in the modelling of multiple $pp$ interactions in the bunch crossing (10\%), and due to MC statistical uncertainties (10\%). To evaluate the dominant 15\% uncertainty due to the modelling of displaced tracks and DV reconstruction, a sample of $K^0_{\mathrm{S}}$ mesons is selected from two-pion vertices in the invariant mass window $488$--$508$~MeV. The rate of $K^0_{\mathrm{S}}$ reconstruction is parameterised as a function of the sum of the $\pt$ of the two tracks ($\sum p_T$) and the radial distance ($r_{DV}$). In each window of $\sum p_T$ and $r_{DV}$, the efficiency is obtained by dividing the event yield by the expected exponentially falling distance distribution, and a weight is computed as the ratio between measured and simulated efficiencies. These weights are normalised to be equal to one at small radii and then used to reweight the DV efficiencies in the signal samples, and a relative difference of 15\% is found as the maximum effect in the final selection efficiency. In addition, uncertainties in the $W$ boson production cross section (3\%) and the integrated luminosity (2.2\%) are taken into account for the interpretation. The total systematic uncertainty, with all contributions added in quadrature, is $24\%$.
 
\subsection{Backgrounds (displaced signature)}
\label{dHNL:backgrounds}
 
Possible background sources which can result in two-track DVs include hadronic interactions in material, decays of metastable particles such as bottom, charm, and strange hadrons, accidental crossings of charged particles produced in the collisions, and cosmic-ray muons which either cross a charged particle from the collision or are reconstructed as two back-to-back muons. All of these are reduced by over an order of magnitude when requiring a prompt muon in the same event, and all except cosmic-ray muons are significantly reduced when requiring both tracks to be matched to objects satisfying tight lepton identification. No backgrounds from single cosmic-ray muons reconstructed as two back-to-back muons remain after applying the cosmic-ray veto described in Section~\ref{dHNL:selection}. Other backgrounds can arise in processes such as dijets and $W$+jets. These are processes with large cross sections combined with a very low probability to produce a DV satisfying the selection criteria, making these backgrounds extremely difficult to reproduce in simulations due to the large numbers of events which need to be simulated. Therefore, they are studied and evaluated in a fully data-driven manner using suitable control regions.
 
A study of the different types of background sources was performed using a control sample of events which fail the requirement of a prompt muon with a matched track in the ID, which is part of the preselection (Section~\ref{dHNL:trigger}). This control sample has undergone LRT and thus consists of events collected by a variety of different triggers and preselections. With the requirement of at least one DV, this sample contains 12 times more events than the sample of events passing the preselection. Sub-categories are defined for DVs containing no reconstructed lepton (0-lepton) and only one reconstructed lepton (1-lepton), as requiring two leptons (2-lepton) would lead to an insufficient number of events for this study. For $m_{\textrm{DV}}<2.5$~GeV, more events with 0-lepton and 1-lepton DVs are observed in the high-density material region as compared to the low-density region (as expected from hadronic interactions in material). Likewise, more opposite-charge events are observed than same-charge events (as expected from metastable particle decays). This shows that these two types of backgrounds contribute at low mass. Above $m_{\textrm{DV}}=2.5$~GeV, the $m_{\textrm{DV}}$ distributions fall quickly and coincide within 5\% regardless of the DV being in a high-density or low-density material region, regardless of the tracks being same-charge or opposite-charge, and regardless of the number of tight leptons identified in the DV. The conclusion of this study is that, in the signal region ($m_{\textrm{DV}}>4$~GeV), backgrounds from hadronic interactions and metastable particle decays provide a contribution of less than 5\% relative to other sources of background.
 
Using the same control sample as above with two muons identified in the DV, peaks in the $m_{\textrm{DV}}$ distribution are observed at the mass value of the $J/\psi$ and $\psi(2S)$ mesons (3.1~GeV and 3.7~GeV, respectively). These peaks correspond to decay distances in the range $4$--$40$~mm, consistent with the decays of $b$-hadrons into $J/\psi$ and $\psi(2S)$. From integrating fits to these two peaks in the region $m_{\textrm{DV}}>4$~GeV, taking into account the fact that this sample is statistically enhanced relative to the samples used in the search as well as fit uncertainties, a contribution of less than 0.005 background events from $J/\psi$ and $\psi(2S)$ decays is estimated in the signal region. No contributions from the very short-lived $\Upsilon$ meson are expected due to the fact that the probability for them to result in displaced decays is negligible, and indeed no two-muon DVs are found for $m_{\textrm{DV}}>3.85$~GeV in the control sample.
 
\begin{table}
\centering
\caption{Observed number of events in the control and validation regions used for evaluating the backgrounds in the displaced HNL signature, including all sources of background except for single cosmic-ray muons reconstructed as back-to-back muons and metastable particles. Transfer factors to translate the number of 0-lepton DVs to 2-lepton DVs are obtained using a same-charge DV control sample. Validation is performed using 1-lepton DVs. The number of events in the signal and validation regions is also indicated. }
\begin{tabular}{lccr}
\toprule
Leptons in DV & Same-charge DV & Opposite-charge DV & Opposite-charge DV estimated\\
\midrule
2           &  0             & 0 (signal region)  & $<2.3$ at $90\%$ CL \\
1 ($\mu$)   &  83            & 89                 & 82.4$\pm 9.0$  \\
1 ($e$)     &  28            & 35                 & 27.8$\pm 5.3$  \\
0           &  169254        & 168037             &   \\
\bottomrule
\end{tabular}
\label{tab:dHNLABCD}
\end{table}
 
The number of background events in the signal region, which requires opposite-charge DVs, is estimated using a control region of same-charge DVs, all other requirements being the same. This is done by applying a transfer factor from 0-lepton DVs to 2-lepton DVs obtained in the same-charge control region. This provides an unbiased estimate of all backgrounds remaining after the selection for which the ratio of 2-lepton background DVs to 0-lepton background DVs does not depend on the DV charge configuration. While this is the case for accidental crossings including those involving cosmic-ray muons, it does not include single cosmic-ray muons reconstructed as back-to-back muons nor decays of neutral hadrons which are either metastable or produced in material interactions, which can be neglected in the $m_{\textrm{DV}}>4$~GeV region as discussed above. The validity of the method is verified by performing the estimate on a validation sample of 1-lepton DVs. The numbers of events observed in the control regions are reported in Table~\ref{tab:dHNLABCD}. The estimated numbers of events in the 1-lepton DV validation region for electron and muon are compatible with the observed numbers within statistical uncertainties. From a 90\% confidence-level limit of 2.3 background events with 2-lepton DVs in the same-charge DV control region (where 0 are observed), an upper limit of $2.3 \times 168037/169254 \sim 2.3$ is obtained for the 2-lepton DVs in the signal region (where 0 are observed).

 
\section{Results}
\label{sec:results}
 
\begin{figure}[tb]
\centering
\ifint
\includegraphics[width=0.99\linewidth]{fig_06a.png}
\includegraphics[width=0.99\linewidth]{fig_06b.png}
\else
\includegraphics[width=0.99\linewidth]{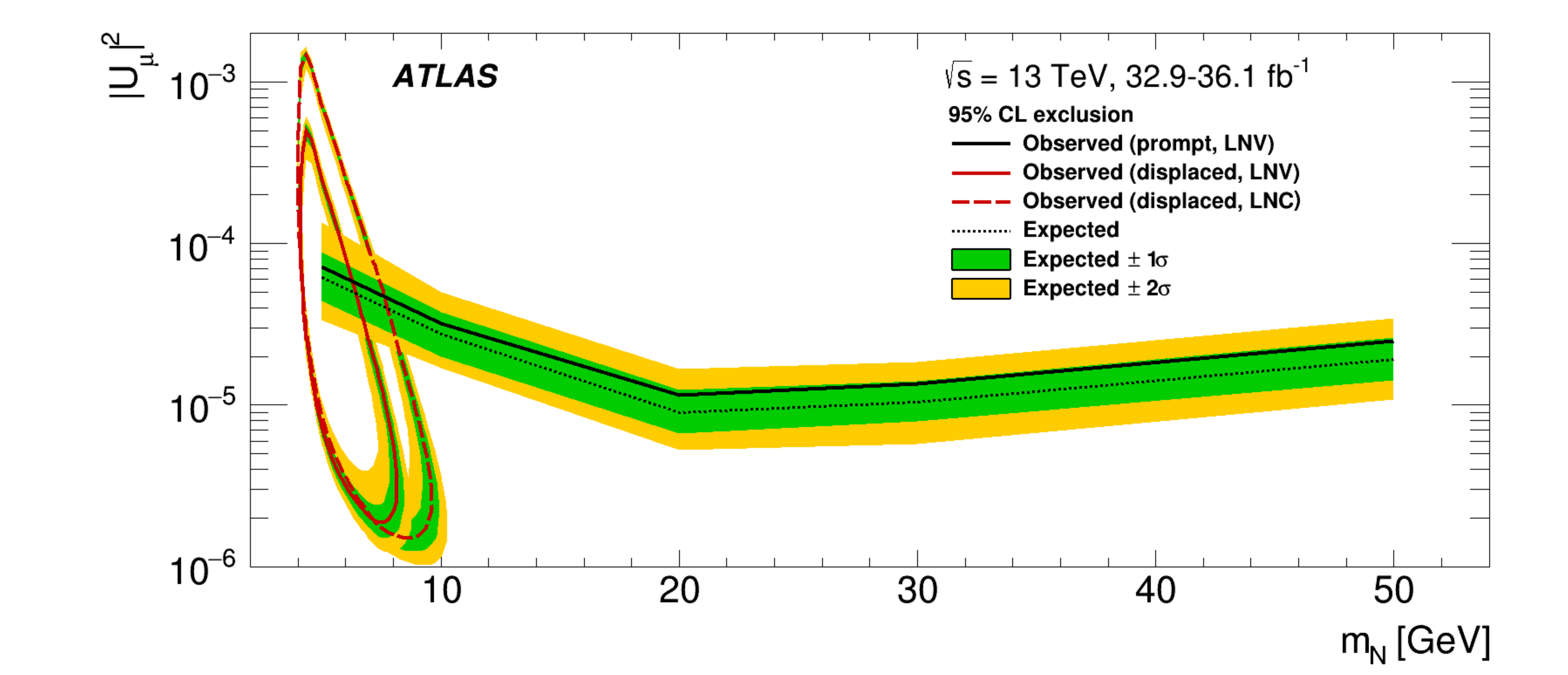}
\includegraphics[width=0.99\linewidth]{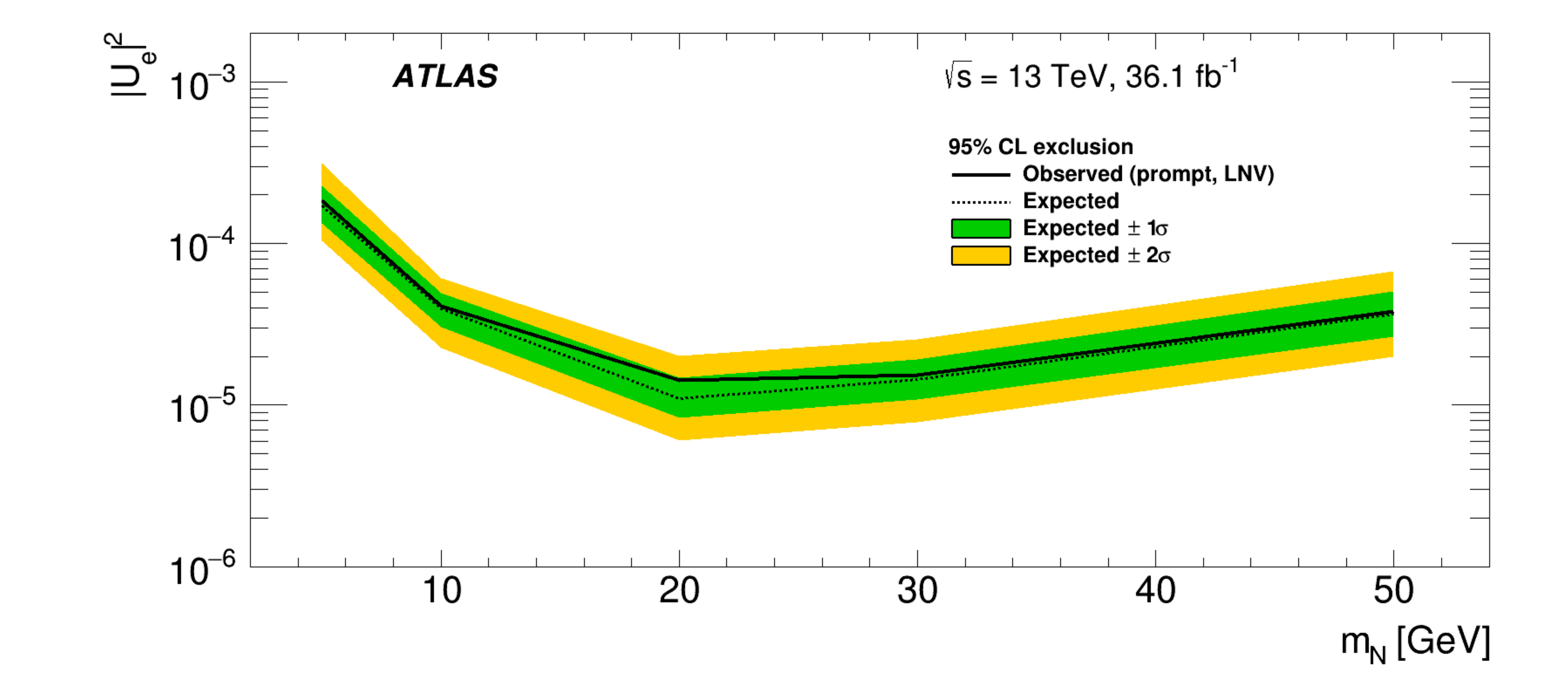}
\fi
\caption{Observed 95\% confidence-level exclusion in $|U_{\mu}|^2$ (top) and $|U_{e}|^2$ (bottom) versus the HNL mass for the prompt signature (the region above the black line is excluded) and the displaced signature (the region enclosed by the red line is excluded). The solid lines show limits assuming lepton-number violation (LNV) for 50\% of the decays and the long-dashed line shows the limit in the case of lepton-number conservation (LNC). The dotted lines show expected limits and the bands indicate the ranges of expected limits obtained within $1\sigma$ and $2\sigma$ of the median limit, reflecting uncertainties in signal and background yields. }
\label{fig:limits}
\end{figure}
 
Observations in the signal regions are consistent with background expectations in both the prompt and displaced signatures described in Sections~\ref{sec:pHNL} and \ref{sec:dHNL}, respectively. For a given HNL mass and lifetime, the selection efficiency is obtained from MC simulations, and the uncertainty in the efficiency is evaluated. Combining this information with the event yield obtained from the integrated luminosity of 36.1~fb$^{-1}$ and 32.9~fb$^{-1}$ for the prompt and displaced analyses, respectively, and the HNL production cross section and branching ratio (Eq.~(\ref{equ:production})), a set of choices of HNL coupling strengths ($|U|^2$) and masses ($m_N$) are excluded at the 95\% confidence level. Calculations of confidence intervals and hypothesis testing are performed using a frequentist method with the CL$_{\mathrm{S}}$ formalism as implemented in RooStats~\cite{Verkerke2003}. The exclusion limits are shown in Figure~\ref{fig:limits}, in the cases of dominant mixing to $\nu_\mu$ (top) and $\nu_e$ (bottom), for the cases of LNV (both signatures, solid lines) and LNC (displaced signature, long-dashed line).
 
Limits from the prompt signature cover the mass range $5$--$50$~GeV. In the mass range $20$--$30$~GeV, the regions in $|U_{\mu}|^2$ and $|U_{e}|^2$ above $1.4\times 10^{-5}$ are excluded, a reach which is limited by the integrated luminosity of the analysed data, as well as the selection efficiency and the signal-to-background ratio. At higher masses, the sensitivity decreases due to a kinematic suppression of HNL production from the $W$ boson decay. For masses below 20~GeV, the long decay path causes large efficiency losses. This is the region where the displaced signature becomes more sensitive.
 
Limits from the displaced signature cover the mass range $4.5$--$10$~GeV, in which they exclude coupling strengths down to $|U_{\mu}|^2\sim 2\times 10^{-6}~(1.5\times 10^{-6})$ assuming LNV (LNC). For comparison, the best previous constraints on $|U_{\mu}|^2$ in this mass range were obtained with the DELPHI experiment at LEP1~\cite{Delphi1997}, excluding values down to $\sim 1.5\times 10^{-5}$. The limit contour of the displaced signature takes the shape of an oblique ellipse which approximately corresponds to HNL proper decay lengths in the range $1$--$30$~mm. It is also limited from below by the product of integrated luminosity and efficiency. The interpretation with LNV provides weaker limits because the search is sensitive to long lifetimes and, for a given coupling strength, the lifetime is reduced by a factor of two when LNV decays are allowed.

 
\newpage
 
\section{Conclusions}
\label{sec:conclusion}
 
\FloatBarrier
 
A search for heavy neutral leptons (HNLs) produced in leptonic decays of on-shell $W$ bosons has been performed using data recorded by the ATLAS detector at the LHC in proton--proton collisions at a centre-of-mass energy of 13~TeV corresponding to an integrated luminosity of up to 36.1~fb$^{-1}$, using two distinct signatures. The prompt signature requires three prompt leptons (either muons or electrons) with no same-flavour opposite-charge configuration. It probes mean HNL proper decay lengths of 1~mm or less, with the assumption of lepton-number violation. The displaced signature, explored for the first time at the LHC, features a prompt muon accompanied by a vertex displaced in the radial direction by $4$--$300$~mm from the beam line containing two opposite-charge leptons (either two muons or a muon and an electron) with a reconstructed vertex mass $m_{\textrm{DV}}>4$~GeV. It does not require lepton-number violation and probes longer lifetimes, corresponding to lower masses.
 
Observations are consistent with background expectations and results of the search are presented as exclusion contours in the HNL coupling strength versus mass plane in a model postulating a single HNL mixing either to muon or electron neutrinos. The prompt signature excludes coupling strengths above $4 \times 10^{-5}$ in the mass range $10$--$50$~GeV, with a most stringent limit of $1.1 \times 10^{-5}$ for a mass of 20~GeV. In the case of mixing to muon neutrinos, the displaced signature excludes coupling strengths down to $2 \times 10^{-6}~(1.5 \times 10^{-6})$ at best in the mass range $4.5$--$10$~GeV assuming lepton-number violation (conservation), surpassing the best previous constraints using on-shell $Z$ boson decays at LEP1 by one order of magnitude.
 
A notable characteristic of the displaced signature is that backgrounds fall off sharply with the vertex mass. Heavy neutral leptons with smaller coupling strengths (longer lifetimes) at higher masses can be probed by increasing the $W$ boson yield at higher luminosities. Also, even though this was not done in this work, it is possible to probe long-lived HNL mixing to the electron flavour, and also to consider their semileptonic decays.
 
\FloatBarrier
 
\section*{Acknowledgements}
 
We thank CERN for the very successful operation of the LHC, as well as the
support staff from our institutions without whom ATLAS could not be
operated efficiently.
 
We acknowledge the support of ANPCyT, Argentina; YerPhI, Armenia; ARC, Australia; BMWFW and FWF, Austria; ANAS, Azerbaijan; SSTC, Belarus; CNPq and FAPESP, Brazil; NSERC, NRC and CFI, Canada; CERN; CONICYT, Chile; CAS, MOST and NSFC, China; COLCIENCIAS, Colombia; MSMT CR, MPO CR and VSC CR, Czech Republic; DNRF and DNSRC, Denmark; IN2P3-CNRS, CEA-DRF/IRFU, France; SRNSFG, Georgia; BMBF, HGF, and MPG, Germany; GSRT, Greece; RGC, Hong Kong SAR, China; ISF and Benoziyo Center, Israel; INFN, Italy; MEXT and JSPS, Japan; CNRST, Morocco; NWO, Netherlands; RCN, Norway; MNiSW and NCN, Poland; FCT, Portugal; MNE/IFA, Romania; MES of Russia and NRC KI, Russian Federation; JINR; MESTD, Serbia; MSSR, Slovakia; ARRS and MIZ\v{S}, Slovenia; DST/NRF, South Africa; MINECO, Spain; SRC and Wallenberg Foundation, Sweden; SERI, SNSF and Cantons of Bern and Geneva, Switzerland; MOST, Taiwan; TAEK, Turkey; STFC, United Kingdom; DOE and NSF, United States of America. In addition, individual groups and members have received support from BCKDF, CANARIE, CRC and Compute Canada, Canada; COST, ERC, ERDF, Horizon 2020, and Marie Sk{\l}odowska-Curie Actions, European Union; Investissements d' Avenir Labex and Idex, ANR, France; DFG and AvH Foundation, Germany; Herakleitos, Thales and Aristeia programmes co-financed by EU-ESF and the Greek NSRF, Greece; BSF-NSF and GIF, Israel; CERCA Programme Generalitat de Catalunya, Spain; The Royal Society and Leverhulme Trust, United Kingdom.
 
The crucial computing support from all WLCG partners is acknowledged gratefully, in particular from CERN, the ATLAS Tier-1 facilities at TRIUMF (Canada), NDGF (Denmark, Norway, Sweden), CC-IN2P3 (France), KIT/GridKA (Germany), INFN-CNAF (Italy), NL-T1 (Netherlands), PIC (Spain), ASGC (Taiwan), RAL (UK) and BNL (USA), the Tier-2 facilities worldwide and large non-WLCG resource providers. Major contributors of computing resources are listed in Ref.~\cite{ATL-GEN-PUB-2016-002}.

\printbibliography
 
\clearpage
 
 
\begin{flushleft}
{\Large The ATLAS Collaboration}

\bigskip

G.~Aad$^\textrm{\scriptsize 102}$,    
B.~Abbott$^\textrm{\scriptsize 129}$,    
D.C.~Abbott$^\textrm{\scriptsize 103}$,    
O.~Abdinov$^\textrm{\scriptsize 13,*}$,    
A.~Abed~Abud$^\textrm{\scriptsize 71a,71b}$,    
K.~Abeling$^\textrm{\scriptsize 53}$,    
D.K.~Abhayasinghe$^\textrm{\scriptsize 94}$,    
S.H.~Abidi$^\textrm{\scriptsize 167}$,    
O.S.~AbouZeid$^\textrm{\scriptsize 40}$,    
N.L.~Abraham$^\textrm{\scriptsize 156}$
% ...
\bigskip
\\

% ...
$^{13}$Institute of Physics, Azerbaijan Academy of Sciences, Baku; Azerbaijan.\\
% ...
$^{40}$Niels Bohr Institute, University of Copenhagen, Copenhagen; Denmark.\\
% ...
$^{53}$II. Physikalisches Institut, Georg-August-Universit\"{a}t G\"ottingen, G\"ottingen; Germany.\\
% ...
$^{71}$$^{(a)}$INFN Sezione di Pavia;$^{(b)}$Dipartimento di Fisica, Universit\`a di Pavia, Pavia; Italy.\\
% ...
$^{94}$Department of Physics, Royal Holloway University of London, Egham; United Kingdom.\\
% ...
$^{102}$CPPM, Aix-Marseille Universit\'e, CNRS/IN2P3, Marseille; France.\\
$^{103}$Department of Physics, University of Massachusetts, Amherst MA; United States of America.\\
% ...
$^{129}$Homer L. Dodge Department of Physics and Astronomy, University of Oklahoma, Norman OK; United States of America.\\
% ...
$^{156}$Department of Physics and Astronomy, University of Sussex, Brighton; United Kingdom.\\
% ...
$^{167}$Department of Physics, University of Toronto, Toronto ON; Canada.\\
% ...
$^{a}$ Also at Borough of Manhattan Community College, City University of New York, New York NY; United States of America.\\
$^{b}$ Also at Centre for High Performance Computing, CSIR Campus, Rosebank, Cape Town; South Africa.\\
% ...
$^{*}$ Deceased

\end{flushleft}


\end{document}